\documentclass[12pt]{article}

\begin{document}
\baselineskip=0.7cm
\newcommand{\EQ}{\begin{equation}}
\newcommand{\EN}{\end{equation}}
\newcommand{\EQA}{\begin{eqnarray}}
\newcommand{\EQN}{\end{eqnarray}}
\newcommand{\EQAN}{\begin{eqnarray*}}
\newcommand{\EQNN}{\end{eqnarray*}}
\newcommand{\e}{{\rm e}}
\newcommand{\Sp}{{\rm Sp}}
\renewcommand{\theequation}{\arabic{section}.\arabic{equation}}
\newcommand{\Tr}{{\rm Tr}}
\newcommand{\lpartial}{\buildrel \leftarrow \over \partial}
\newcommand{\rpartial}{\buildrel \rightarrow \over 
\partial}
\renewcommand{\thesection}{\arabic{section}.}
\renewcommand{\thesubsection}{\arabic{section}.\arabic{subsection}}
\makeatletter
\def\section{\@startsection{section}{1}{\z@}{-3.5ex plus -1ex minus 
 -.2ex}{2.3ex plus .2ex}{\large}} 
\def\subsection{\@startsection{subsection}{2}{\z@}{-3.25ex plus -1ex minus 
 -.2ex}{1.5ex plus .2ex}{\normalsize\it}}
\makeatother
\def\thefootnote{\fnsymbol{footnote}}

\begin{flushright}
hep-th/0308024\\
UT-KOMABA/03-16\\
KEK-TH-905\\
July 2003
\end{flushright}
\vspace{0.3cm}
\begin{center}
\Large
PP-Wave Holography for D{\it p}-Brane Backgrounds
\vspace{0.7cm}

\normalsize
Masako  {\sc Asano},  
\footnote{
e-mail address:\ \ {\tt  asano@post.kek.jp}}
\qquad 
 Yasuhiro {\sc Sekino}, 
\footnote{
e-mail address:\ \ {\tt  sekino@post.kek.jp}}
\vspace{0.2cm}

{\it Theory Division, Institute of Particle and Nuclear 
Studies KEK\\High Energy Accelerator Research 
Organization, Tsukuba, Ibaraki 305-0801}

\vspace{0.3cm}
and

\vspace{0.3cm}
Tamiaki {\sc  Yoneya}
\footnote{
e-mail address:\ \ {\tt tam@hep1.c.u-tokyo.ac.jp}}
\\
\vspace{0.2cm}

{\it Institute of Physics, University of Tokyo\\
Komaba, Meguro-ku, Tokyo 153-8902}

\vspace{0.6cm}
Abstract
\end{center}
As an extension of the so-called BMN 
conjecture, 
we investigate the plane-wave limit for
 possible holographic connection 
between bulk string theories in 
non-conformal backgrounds of D$p$-branes and the corresponding supersymmetric 
gauge theories for $p<5$. Our work 
 is based on the 
tunneling picture for dominant null trajectories 
of strings in  the limit 
of large angular momentum. 
The tunneling null trajectories start from the near-horizon boundary and return to the boundary, 
and the resulting backgrounds are 
time-dependent for general D$p$-branes 
except for $p=3$. We develop a general method for extracting diagonalized two-point functions 
for boundary theories as Euclidean (bulk) S-matrix in the  time-dependent backgrounds.  
For the case of D0-brane, 
two-point functions of supergravity modes 
are shown to 
agree with the results derived previously 
by the perturbative analysis of supergravity. 
We then discuss the implications 
of the holography for general cases of D$p$-branes 
including the stringy excitations. All the cases 
($p\ne 3, p<5$) exhibit 
interesting infra-red behaviors, which are different 
from free-field theories, suggesting the existence of 
quite nontrivial fixed-points in dual gauge theories.

\newpage
\section{Introduction}
One of remarkable 
 developments in string theory in recent years has been 
the formulation of `holographic' connection 
between string theories in bulk space-times and 
supersymmetric gauge theories defined on their 
boundaries. The most typical and best known case 
is the AdS/CFT correspondence between 
type IIB string theory around the large number 
($N$) of D3-branes and maximally supersymmetric 
${\cal N}=4$ Yang-Mills  theory in four dimensions in the 
large $N$ strong coupling  limit $R^{4}=g^2_{{\rm YM}}N 
\rightarrow \infty$. This case is very special 
in the sense that the system is governed by 
exact superconformal symmetry 
on both sides of bulk and boundary space-times, 
which is quite restrictive in 
constraining the dynamics of the systems. 
There are many reasons to expect that the (super)conformal symmetry is not prerequisite for such holographic connections. 
In the case of general D$p$-branes, there exists no 
conformal symmetry and hence the comparison of 
both sides necessarily requires much more 
deeper understanding about dynamics 
 especially on the side of gauge-theories. 
On the side of bulk string theories, 
what we need  is the study of propagating closed strings 
among source-and-probe D$p$-branes. In supergravity 
approximation, this is in principle straightforward, 
though full stringy treatment is in general very 
hard, owing to the complicated structure of 
background space-times. However, once we have 
definite results on  the bulk side, we would have 
predictions for the dynamics of large $N$ strong ($g_{{\rm YM}}^2N \gg 1$) coupling 
gauge theories from closed string theories. 

        From this point of view, an important step 
is the proposal made in 
ref. \cite{bmn} concerning the gauge 
invariant operators (BMN operators) of the 
${\cal N}=4$ super Yang-Mills theory which correspond to the higher 
stringy modes in the bulk for a special class of states with  infinitely large orbital angular momentum $J\sim R^2$. 
The effect of stringy excitations 
is reflected to anomalous 
conformal dimensions 
of the (non BPS) BMN operators, which
 are predicted to behave as   $\sqrt{1+{R^4n^2\over J^2}}$, since 
the world-sheet string theory in the same limit is 
described as a free massive field theory with 
mass of order $J^2/R^4$, as derived by taking the so-called Penrose limit \cite{penrose} 
around the null geodesics describing the 
trajectories of strings in the limit $R^2 \sim J \rightarrow 
\infty$.\footnote{
In the present paper, we always use the unit $\sqrt{\alpha'}=\ell_s=1$. }   
This prediction is 
based on the usual assumption that the energy 
with respect to global time coordinate of AdS geometry 
corresponds to conformal dimension of CFT. 
If this is correct,  the conformal dimension $\sqrt{1+{R^4n^2\over J^2}}=1+ {R^4n^2\over 2J^2} +
\cdots $ for stringy operators 
($n\ne 0$) has a smooth analytic 
behavior with respect 
to the `t Hooft coupling $g^2_{{\rm YM}}N\sim R^4$, 
and hence is accessible by perturbative computations  
on the gauge theory side. 
A number of explicit 
computations using perturbation theory have 
been reported on the gauge-theory side, and 
consistency with the above predictions has been largely 
confirmed, although 
some crucial issues related to the interpretation of  holography and to the derivation of correlation functions 
are being in controversy. For a (partial)  list of works 
on this subject,  we refer the reader to  
reviews appeared recently \cite{rev}.

It is tempting to extend this 
proposal to a more general case of D$p$-branes 
\cite{itzhaki} 
and to see whether or not similar behaviors can be 
expected for nonconformal cases. 
In particular, for D0-branes, that would give some 
important predictions on the behavior of 
the corresponding gauge theory, namely, M(atrix) theory 
in a large $N$ strong coupling limit, with 
respect to the stringy degrees of freedom in 
terms of the matrix variables.  In M(atrix) theory, 
a simple perturbative analysis on the gauge theory side 
cannot give reliable results on the large $N$ 
strong coupling behavior, 
because of severe infrared 
problems. Even two-point functions can have 
very nontrivial behavior in the absence of 
conformal symmetry. 
On the other hand, it is well known that  
the Penrose limit \cite{blauetal} of general D$p$-brane 
backgrounds leads to 
world-sheet theories with intrinsically time-dependent masses except 
for $p=3$~\cite{gimonetal}. Thus, the absence of conformal symmetry 
corresponds to the time dependence, and 
 one of the main problems to be 
overcome in extending the 
BMN conjecture boils down to 
extracting definite predictions from 
string theories with time dependent mass terms. 
This is by itself an interesting question as an example 
in extending the usual treatments of string theory in simple 
time-independent backgrounds to 
time-dependent cases. 
To our knowledge, from the viewpoint of holography, no concrete results have been reported 
in the literature along this line: 
We do not know 
how the correspondence between 
conformal dimensions of 
boundary theory and light-cone energy 
of bulk theory 
should be generalized to nonconformal 
and time-dependent 
world sheet theories, respectively, for general D$p$-branes.  
Actually, even in the typical case of 
D$3$-branes, the usual treatments do not give 
any definite prescription on how the two-point 
functions of boundary theory is computed directly 
from the bulk theory, in the absence of 
concrete holographic dictionaries between bulk and 
boundary theories.

In fact, for general D$p$-branes there is a 
pseudo-symmetry called the `generalized conformal 
symmetry', as proposed in \cite{jeyone}. In the case 
of D0-branes
for definiteness, 
the generalized 
conformal symmetry combined with 
some natural assumptions related to 
holography predicts \cite{yopotsdam} that 
two-point functions for at least supergravity 
states and the corresponding gauge-theory 
operators have a general form 
\EQ
\langle{\cal O}_I(t_1){\cal O}_I(t_2)\rangle
\sim {1\over g_s^2\ell_s^8}
(g_sN\ell_s^7)^{(\Delta_I+6)/5} |t_1-t_2|^{-(7\Delta_I+12)/5}
\EN
under the following assumptions:
\begin{enumerate}
\item They are diagonalized,  
and the operators ${\cal O}_I$ have 
definite scaling dimensions in the sense that 
\EQ
{\cal O}_I(t) \rightarrow
{\cal O}_I'(t') =\lambda^{\Delta_I}{\cal O}_I(t),
\quad t\rightarrow t'=\lambda^{-1}t, \quad
g_s\rightarrow g_s'=\lambda^3 g_s. 
\label{genescaling}
\EN
The operators are normalized such that their engineering 
dimensions with respect to length are equal to $-1$. 
\item The two-point correlation functions 
should be inversely proportional to 10 dimensional Newton 
constant $G_{10}\sim g_s^2\ell_s^8$ and, apart 
from this prefactor,   the {\it only} possible 
parameter entering in the two-point functions is 
the length scale 
$(g_sN\ell_s^7)^{1/7}$. 
\end{enumerate}
The second assumption comes from the  basic holographic relation 
that two-point functions are obtained 
by diagonalizing 
linearized fluctuations in the bulk supergravity fields 
around the D0-brane background, as reviewed briefly in 
Appendix A. 
Recall that the gravitational 
potential around $N$ D0's is $(g_sN\ell_s^7)/r^7$, which 
implies that the near-horizon dynamics should be 
governed by the length scale $(g_sN)^{1/7}\ell_s$.  
Note that this scale is different from the 
{\it elementary} D0-scale $\ell_M\sim g_s^{1/3}\ell_s$ 
which is nothing but the M-theory scale. 
This assumption demands that the 
two-point functions take the 
form $G_{10}^{-1}f((g_sN\ell_s^7)^{1/7}, |t-t'|)$.
The transformation 
rule in (\ref{genescaling}) has its origin in the invariance of 
the M(atrix) theory action and also of the D0-background 
under the transformation 
$X(t)\rightarrow X'(t') =\lambda X(t), \quad t'=\lambda^{-1}t, \quad
g_s\rightarrow g_s'=\lambda^3 g_s$, which is motivated by the space-time uncertainty relation (see ref. \cite{yoptp}), $\Delta t \Delta X \ge 
\ell_s^2 $, and 
is equivalent to the kinematical Lorentz boost along the 
M-theory direction.   
The power-law behavior satisfied by this prediction
 reflects the 
absence of mass gap in this system, as it should be 
in any theory of gravity. 
The spectrum for the 
dimensions $\{\Delta_I\}$ has been derived 
in previous works  \cite{sekiyone}, 
by performing detailed supergravity analyses 
which have confirmed the above prediction. 
The results, 
$\Delta_I =4\ell_I/7 +2n-3$ (or $\Delta_I =4\ell_I/7 +2n 
-3/2$ for 
fermionic operators) with $\ell_I$ being the angular momentum and $n$ being non-negative 
integers,  are consistent with the behavior 
of M(atrix) theory operators 
known approximately  from one-loop 
analysis in \cite{kabataylor}, with some slight but 
puzzling corrections as discussed in \cite{yopotsdam, sekiyone}.
In view of this, it 
seems reasonable to expect that certain 
appropriate extensions of the BMN correspondence 
exist for non-conformal case of  D$p$-branes. 

The purpose of the present paper is to provide a first step 
along this line. We develop a general method 
of extracting diagonalized two-point functions 
at the boundary as the Euclidean S-matrix in 
the bulk. 
Based on this general method, we derive the 
spectrum of $\{\Delta_I \}$, confirming and 
generalizing the results of the previous supergravity analyses 
at least in the case of bosonic excitations.  
As we will see, holography predicts that 
the behaviors of stringy BMN 
operators in nonconformal cases 
are in general quite different from the 
conformal case of D$3$-branes. 

 For performing the required analysis  in a clear 
space-time picture, it is very crucial to 
base our discussion of the Penrose limit 
on the tunneling picture, as proposed in 
refs. \cite{dobashimayone} \cite{yonishiyukawa} for the case of D3-branes, 
which allows us to avoid singularities 
at the horizon for 
nonconformal D$p$-branes and to establish a direct connection 
between bulk and boundary. In fact, to our knowledge, 
there has been no other proposals which have given  
 concrete prescriptions  
on the {\it direct} 
relationship between bulk amplitudes 
and boundary correlators in such a general 
way as allowing extensions to non-conformal cases.   
Therefore, in the next 
section, we begin by recapitulating 
main points of this proposal 
adapted to the general 
D$p$-branes. Namely, we argue that,  in the limit 
of large angular momentum $J$, two-point 
functions in dual gauge theories should be 
described by transition amplitudes 
(which we call collectively the {\it Euclidean} S-matrix) 
defined along tunneling null trajectories 
traversing from boundary to boundary in the bulk. 
Then we are 
led to world-sheet theories with time dependent 
backgrounds, since the effective masses 
of strings are in general not constant along the 
trajectories. 
 A general quantum theory with time-dependent 
mass is given in section 3. The discussion 
is sufficiently general and 
might be interesting in its own right apart 
from our specific application 
treated in the present work. In section 4, 
we derive diagonalized 
two-point functions for D$p$-branes on the basis 
of the general formalism of section 3. We 
compare the results with 
previous supergravity analysis  in the case 
of D0-branes and further 
consider the cases of general D$p$-branes 
 including stringy excitations.   In particular, we 
discuss the predictions from holography on the 
infra-red limit of the dual gauge theories of 
D$p$-branes.  
In section 5, 
we conclude by summarizing the present work 
and by mentioning 
possible future directions. 
We also point out 
very interesting possible  implications from our results on the infra-red behaviors for the cases $p=1$ and $p=4$, 
suggesting the shifts 
$d=p+1\rightarrow d_{{\rm eff}}=d+1$ of effective dimensionalities. 
A  brief summary of field-theory analyses  in supergravity 
\cite{sekiyone} with partial extensions to 
general case of D$p$-brane backgrounds are given 
in Appendix A for the purpose of making the present 
exposition reasonably self-contained. 
Appendix B is devoted to a side remark on an alternative 
effective PP-wave description of the supergravity 
fluctuations in the case of D0 backgrounds. 
This description is useful to confirm 
our results in the case of non-stringy supergravity 
modes from a perspective which is 
slightly different from the main text.  

\section{PP-wave holography and tunneling null 
geodesics}
\setcounter{equation}{0}

\subsection{Null geodesics in the D$p$-brane backgrounds}
As is well known, the Penrose limit can be 
regarded as  a semiclassical limit \cite{gkp2}\cite{tseytlin} 
for the 
propagation of particles or strings along 
null geodesics with large (angular) momentum. 
Let us recall the classical metric around $N$ D$p$-branes 
in the near horizon limit  $(-\pi/2 \le \psi \le \pi/2)$
\EQ
ds^{2}=q_p^{1/2}
\Big[H^{-1/2}(-dt^{2}+d\tilde{x}_{a}^{2})
+H^{1/2}(dr^{2}+r^{2}d\psi^{2}+r^{2}\cos^{2}\psi 
d\Omega_{7-p}^{2})\Big] 
,
\label{Dpmetric}
\EN 
where $H=1/r^{7-p}$ and $q_p=2^{7-2p}\pi^{(9-3p)/2}\Gamma((7-p)/2)
g_{{\rm YM}}^2N$.
We have rescaled the coordinates along the D-branes $(t, \tilde{x}_{a})
\rightarrow q_p^{1/2}(t, \tilde{x}_{a})$ $(a=1,\ldots, p)$
from the usual convention, such that the characteristic dimensional parameter $q_p^{1/2}$ appears as the overall 
prefactor in the metric. 
The metric can also be written in the form
\EQ
ds^{2}=q_{p}^{1/2}e^{2\tilde{\phi}/(7-p)}
\left\{ \left({2\over 5-p}\right)^{2}
{-dt^{2}+d\tilde{x}_{a}^{2}+dz^{2}\over z^{2}}
+d\psi^{2}+\cos^{2}\psi d\Omega_{7-p}^{2}\right\},
\label{eq:Dpmetric2}
\EN
\EQ
e^{\tilde{\phi}}\equiv H^{(3-p)/4}=\left( {5-p\over 2}z
\right)^{(7-p)(3-p)\over 2(5-p)}.
\label{etildephi}
\EN
The radial coordinate $z$ is defined by
\EQ
z={2\over 5-p}\sqrt{1\over r^{5-p}}
. 
\EN
For $p=3$, (\ref{eq:Dpmetric2}) is the 
${\rm AdS}_{5}\times {\rm S}^{5}$ metric  
for the Poincar\'{e} patch. The dilaton background
is given by $e^{\phi}=g_{s}q_{p}^{(3-p)/4}e^{\tilde{\phi}}$.
We can think of the Weyl factors $q_{p}^{1/2}e^{2\tilde{\phi}/(7-p)}$
 as defining  position-dependent 
effective length scales  for these backgrounds. 

The null trajectories which have nonzero angular 
momentum $J\equiv  EH^{1/2}r^2\dot{\psi}$ for
 the direction of the 
angle $\psi$ and traverses along the radial direction 
of $r$ satisfy 
\EQ
\dot{r}^2=H^{-1}(-\ell^{2}r^{-2}+H)
,   
\label{realtraj}
\EN
where $\ell\equiv J/E$ and $\dot{r}$ is the velocity with respect to 
the affine time $\tau$ satisfying 
the relation
\EQ
EH^{-1/2}\dot{t}=E \rightarrow \dot{t}=H^{1/2}. 
\EN
The parameter $E$ introduced here 
can be interpreted as energy 
with respect to the target time $t$. 
These conventions 
for the target energy and angular momentum are 
appropriate if the particle theory is understood as the 
limit from the string theory as formulated in later 
subsections.  
Providing that we  
restrict ourselves to the case $p<5$, the allowed region is 
\EQ
r\le r_0, \quad r_0=\ell^{-2/(5-p)}.
%\left({1\over \ell^{2}}\right)^{1/(5-p)}. 
\EN
Namely, the null trajectories never reach the 
near-horizon boundary 
$r\sim q_p^{1/(7-p)} $
%$r\rightarrow q_p^{1/(7-p)} $ 
for large $g_sN$  and fall to the horizon at 
$r=0$ which is a singular point except for $p=3$. 
In particular, the trajectories always go to the horizon 
in a finite affine time.  This is problematical 
for at least two reasons. First, hitting singularity 
invalidates the classical approximation itself, and 
secondly, the separation of the trajectories 
from the boundary 
makes holographic correspondence very obscure, 
since the identification of the bulk modes and 
the operators on the boundary is based 
on the behavior of the former near the boundary. 
For $p=3$, we can extend the trajectories without 
encountering singularities to globally defined 
AdS space-time. 
%However this seems to make the 
%second problem worth, 
However, this does not solve the second problem
since it is difficult to 
associate global coordinate to physical events  on the 
boundary.
%and hence holography becomes very obscure. 
For $p=6$ the situation is opposite: The allowed region is 
$r\ge \ell^{2/(p-5)}$. The case $p=5$ is marginal: There is 
no trajectory, either real or tunneling, which 
traverses  from boundary to boundary. In the present paper, we 
 restrict our considerations to $p<5$. 

\subsection{Tunneling null geodesics and holography for D3-branes}
To explain the problem in a simplest possible setting, 
let us for the
moment specialize ourselves to the conformal case $p=3$. 
 The null geodesic is given by
\EQ
z={\ell\over \cos \ell\tau}, \quad t=\ell\tan \ell\tau ,  \quad 
\psi=\ell\tau.
\label{tunnelingtraj} 
\EN
 Since it is easy to check that $\ell\tau$ is 
nothing but the time coordinate of the 
global AdS metric, periodicity with respect to $\tau$ reflects the  fact that the global AdS space-time  
is a universal cover of hyperboloid with topology $S^1\times 
${\bf R}$^{4}$. 
Because this periodicity would affect 
all physical amplitudes defined around this 
background, it seems very difficult to 
adopt the usual interpretation of the affine 
time as proposed in 
\cite{bmn} that $\tau$ should be identified with 
radial time of the boundary theory 
as $\vec{x} \sim e^{\tau}\hat{\vec{x}}$  
$(\vec{x}\equiv (\tilde{x}_a, t)) $. 
It is hard to imagine any periodicity 
in the dynamics of super Yang-Mills theory 
in radial quantization 
on the boundary. In fact, the 
only clue for the holography in terms of physical 
amplitudes in the 
AdS/CFT correspondence is the famous 
GKP-Witten relation \cite{holography} between 
from-boundary-to-boundary amplitudes 
in the bulk space-times and the correlation 
functions of super Yang-Mills theory defined at 
the boundary. The correspondence between 
states $\{\phi^i_0(\vec{x})\}$ at the boundary and the fluctuating fields $\{\phi^i(z, \vec{x})\}$ 
in the bulk is based on the boundary condition
\EQ
\lim_{z\rightarrow 0} \phi^i(z, \vec{x})= z^{4-\Delta_i}
\phi^i_0(\vec{x}) .
\label{boundc}
\EN
The states $\phi^i_0(\vec{x})$ couple to a set of 
local gauge-invariant 
operators $\{{\cal O}_i(\vec{x})\}$
 with definite conformal 
dimensions $\Delta_i$ at the boundary.  Clearly, it is impossible to directly use this correspondence for the propagation of strings along the above trajectory. 
  From the viewpoint at the boundary, the relevant region 
of affine time  seems to be only a finite segment 
$-\pi/2 <\ell \tau<\pi/2$ corresponding to 
$-\infty < t <\infty$. 

The proposal made in ref. \cite{dobashimayone} 
in order to resolve this difficulty  
is simply to consider a complex trajectory 
which represents a tunneling in the semi-classical 
picture from boundary to boundary, instead of the real trajectory going form horizon to horizon.  
Formally, this corresponds to using  
purely imaginary affine time, $\tau\rightarrow -i\tau$,  
  and 
also to Wick-rotating the target time and angles,  $t\rightarrow -it$ and $\psi
\rightarrow -i\psi$. Thus the tunneling trajectory 
is now given by   
\EQ
z={\ell\over \cosh \ell\tau}, \quad t=\ell \tanh \ell\tau ,  \quad 
\psi=\ell\tau. 
\label{D3tunnel}
\EN
This  drastically changes  the global structure of the 
trajectory such that it never reaches the horizon.  
It now traverses from boundary 
($\tau\rightarrow 
-\infty$) to boundary ($\tau\rightarrow \infty$) with an 
infinite time interval. In this case, we cannot 
identify the affine time $\tau$ with 
the radial time of boundary theory, since 
near the boundary the tunneling 
trajectory is {\it orthogonal} to 
D3-brane target space-time. Instead, it must be 
identified with the coordinate $z$ by 
$z \sim e^{-\ell|\tau|}$ near the boundary. 
This is consistent with the identification of 
energy with respect to $\tau$ as 
the  conformal dimensions at the boundary, 
since it is well known that 
the variable $z$ plays the role of effective cutoff 
parameter for the short distance structure of the 
dual gauge theory at the boundary.  It is also important 
that the parameter $2\ell$ is now directly 
interpreted, from the second in the equations 
(\ref{tunnelingtraj}), as the distance of two end-points,  
corresponding to the insertion of 
operators for the boundary theory. 

We emphasize that this change of 
trajectory  solves 
another obvious puzzle which has been 
ignored in the recent literature. The conjecture 
in \cite{bmn} assumes that the transverse directions 
to the null propagation consist of 8 directions, 
of which 4 are directions of $S^5$ orthogonal to the $\psi$-direction and 
of which the remaining 4 are nothing but the 
directions of the world-volume 
of 
D3-branes. The latter fact is manifested in 
the derivatives $D_iZ \, \, (i=1, \ldots, 4)$, which are 
identified as 
the latter 4 transverse degrees of freedom, 
expressed in terms of the complex 
field $Z=\phi_5 + i\phi_6$ corresponding to the 
U(1) R-charge directions.
However, since the affine-time direction along the 
world-sheet (or world line) must be 
orthogonal to the  transverse excitations 
of strings, we cannot think of any directions 
of the base space of the boundary theory as affine-time $\tau$, including a rather familiar identification 
of global time $\ell\tau$ with the time 
of radial quantization of Yang-Mills theories. 
Clearly, the appearance of the derivatives 
along the base-space directions representing 
transverse excitations becomes a contradiction 
if one insists that 
the affine-time flows along any one 
direction of the D3-brane world volume. 
The fact that the Hamiltonian with respect to the 
affine time is essentially the dilatation operator 
at the boundary 
should be interpreted as the (gravitational) 
Hamiltonian constraint describing the {\it dynamics} of this theory, and should be discriminated from the unjustified {\it kinematical}  identification of the affine time with the radial time.

Another related remark at this juncture is that, 
because of the  drastic change of the 
global structure of trajectories,  connecting  two  
pictures, real propagation or complex tunneling,  
is not straightforward, though the 
analytic continuation {\it with respect to 
the target time} $t$ is of course allowed after 
computing  correlators.   For example, 
we would encounter integrations over the 
affine time in computing various physical 
amplitudes. However, we would not be allowed to deform the integration contour 
to real affine time, since we would then  have 
divergent results in general 
because of the periodicity as warned above, 
obstructing the deformation of the integration contours. 
For further discussions related to 
this problem, see \cite{yoneprep}. 

Though the main subject of the present paper is the
the semi-classical quantization of the fluctuations around
tunneling null geodesics and the derivation of 
transition amplitudes for them, it is useful here to 
discuss the two-point amplitude in the classical limit 
taking the example of a massless point particle
on the D3-brane background.
The part of the action containing $(t,z,\psi)$ is
\begin{eqnarray}
S&=&{1\over 2}\int d\tau {1\over \eta} g_{\mu\nu}\dot{x}^{\mu}\dot{x}^{\nu}
\nonumber\\ 
&=&{1\over 2}\int d\tau {1\over \eta}\left\{ {R^{2}\over z^{2}}
(\dot{t}^{2}+\dot{z}^{2})-R^{2}\dot{\psi}^{2}\right\}
\label{eq:particle}
\end{eqnarray}
where $R=q_{3}^{1/4}$ is the common scales of both AdS$_5$ 
and S$^5$, and 
we have kept the einbein $\eta$ unfixed.  
For representing the amplitude for 
a fixed angular momentum \begin{equation}
J={\delta S\over \delta \dot{\psi}}=-R^{2}{\dot{\psi}\over \eta},
\end{equation}
we perform a Legendre transformation to 
Routh function 
\begin{eqnarray}
\bar{S}&=&S-\int d\tau P_{\psi}\dot{\psi}\nonumber\\
&=&{1\over 2}\int d\tau \{{1\over \eta}{R^{2}\over z^{2}}
(\dot{t}^{2}+\dot{z}^{2})+{\eta\over R^{2}}J^{2}\}.
\end{eqnarray}
Solving the equation of motion, taking the null-constraint into account,  and substituting the 
result to the Routh function, we obtain 
\begin{equation}
\bar{S}={J\over R}\int d\tau 
\sqrt{{{R}^{2}\over z^{2}}(\dot{t}^{2}+\dot{z}^{2})}.
\end{equation}
Note that use of the Routh function corresponds to a Fourier
transformation of the wave function 
to the $J$-basis.\footnote{For related treatments of D-particles using Routh functions, see \cite{polchinoy}.}
Thus the amplitude between states with definite $(t,z,J)$
is given by
\EQ\langle t_f,1/\Lambda,J ;T| t_i,1/\Lambda, J;-T\rangle = \int {\cal D}t{\cal D}z e^{-\overline{S}[t,z,J]} 
.
\EN
In the classical approximation,   the result is therefore  
\begin{eqnarray}
 \langle t_f,1/\Lambda,J ;T| t_i,1/\Lambda, J;-T\rangle_{class}&=&e^{-\bar{S}}
%\nonumber\\
=e^{-2J\ell T}=\left( 1 \over \Lambda |t_{f}-t_{i}|\right)^{2J},
\end{eqnarray}
where the last equality is derived as follows: 
From (\ref{D3tunnel}), we see that the distance of two
boundary points where the trajectory begins and ends is $|t_{f}-t_{i}|=2\ell$.
It also follows that the 
cutoff $\Lambda$ for the radial coordinate 
($1/\Lambda\le z \le z_{0}=\ell$) and
the time interval $T$ with respect to $\tau$
($-T\le \tau \le T$) are related as
\begin{equation}
{1\over \Lambda}=2\ell e^{-\ell T}.
\label{eq:param2}
\end{equation}
 Thus apart from the cutoff-dependent 
factor which should be renormalized as has been 
already exhibited in the 
boundary condition (\ref{boundc}), 
the two-point amplitude reduces 
to the expected form of the 2-pt correlator 
for an operator of conformal dimension $\Delta 
\sim J$ for large $J \sim R^2$. 
  As discussed in \cite{dobashimayone} in the WKB approximation, inclusion of linearized fluctuations 
leads to the shift $J\rightarrow J+4$, which is 
appropriate for a scalar operator, whose origin in the present context is the zero-point 
energies, $4=8\times 1/2$, of 8 transverse directions in the 
10D space-time.  

\subsection{Tunneling null geodesics in the D$p$-brane backgrounds}
We now extend the above discussion to general D$p$-branes 
($p<5$).
%Let us return to the case of general D$p$-branes. 
Because of the singularities at the horizon, it is 
crucial to base our discussions on the tunneling 
picture. After the formal Wick rotations ($\tau\to -i\tau$,
$t\to -it$, $\psi\to -i\psi$), 
 the equation determining the trajectories becomes
\EQ
\dot{r}=\pm H^{-1/2}\sqrt{\ell^{2}r^{-2}-H}
=\pm  r_{0}^{-(5-p)/2}\sqrt{r^{5-p}-r_{0}^{5-p}}.
\label{eq:drdtau}
\EN
The allowed region is  $r\ge r_0$ for $p<5$. 
%Finally, we calculate the classical amplitude for the 
%case of D0-brane. 
Comparing with the conformal case 
of D3-branes, there is a notable difference
when $p<3$. To see this, let us examine the equations for 
the trajectory using the $z$ coordinate
\EQ
\dot{z}= \mp \left( {5-p\over 2}\right)^{2}
e^{-{2\over (7-p)}\tilde{\phi}}
z\sqrt{z_{0}^{2}- z^{2}}, 
\quad
\dot{t}= \left( {5-p\over 2}\right)^{2}
z^{2} e^{-{2\over (7-p)}\tilde{\phi}},
\quad z_{0}={2\ell\over 5-p}
,
\EN
where $e^{\tilde{\phi}}$ is
given in (\ref{etildephi}). 
This shows that, when $p<3$, the trajectory 
starting from the 
asymptotic boundary returns to the boundary 
in a {\it finite} affine time $2T_b$  
\EQ
T_b=\left({2\over 5-p}\right)^{7-p\over 5-p} 
\int^{z_{0}}_{0} dz 
{z^{-{2\over 5-p}}\over \sqrt{z_{0}^{2}-z^{2}}}. 
\EN  
The corresponding (coordinate) 
distance between two end-points 
at the boundary is 
\EQ
|t_{f}-t_{i}|=\int^{T_{b}}_{-T_{b}}d\tau \dot{t}
 =2\int_{0}^{z_{0}}dz 
{z\over \sqrt{z_{0}^{2}-z^{2}}}=2z_0={4\over (5-p)}\ell.
\label{eq:tfti}
\EN

We have to set a cut-off in approaching the boundary, 
such that 
$1/\Lambda\le z \le z_{0}$. 
The reason is that, 
in order for the near-horizon approximation to be valid, 
$1/\Lambda$ is assumed to be of order 
$q_p^{-(5-p)/2(7-p)} \, \, (\rightarrow 0$ as 
$g_sN\rightarrow \infty$) 
corresponding to 
$r <q_p^{1/(7-p)}$,
 
as discussed in detail in \cite{sekiyone} for the case of D0-branes. 
The affine time interval is $-T\le \tau \le T$ 
(with $z(\pm T)=1/\Lambda$)
where 
\EQ
T_{b}-T=\left(2\over 5-p\right)^{{7-p\over 5-p}} 
\int^{1/\Lambda}_{0} dz 
{z^{-{2\over 5-p}}\over \sqrt{z_{0}^{2}-z^{2}}}\sim 
{5-p\over 3-p}\left({2\over 5-p}\right)^{{7-p\over 5-p}}
{1\over z_{0}}\Lambda^{-{3-p\over 5-p}}.
\EN
We see below that the two-point amplitudes in general exhibit similar nonanalyticity appeared here 
with respect to the cutoff $\Lambda$.

Following the procedure explained for 
the case of D3-branes, we evaluate the 
Routh function for the tunneling 
null trajectory
in the point-particle case as 
\begin{eqnarray}
\bar{S}&=&S-\int d\tau J\dot{\psi}\nonumber\\
&=&{q_p^{1/2}\over 2}\int d\tau \left\{
{1\over \eta}
%e^{-{2\over (7-p)}\tilde{\phi}}
e^{-2\tilde{\phi}/(7-p)}
{1\over z^{2}}
(\dot{t}^{2}+\dot{z}^{2})
+\left({2\over 5-p}\right)^{2}q_p^{-1}\eta 
%e^{-{2\over (7-p)}\tilde{\phi}}
e^{-2\tilde{\phi}/(7-p)}
J^{2}
\right\}.
\end{eqnarray}
Substituting the classical solution, we find 
\begin{eqnarray}
\bar{S}&=&\left({2\over 5-p}\right)^{2}{J^{2}\over E} \int_{-T}^{T}d\tau
e^{-{2\over (7-p)}\tilde{\phi}}
={4\over 5-p}Jz_{0}\int^{z_{0}}_{1/\Lambda}
dz{1\over z\sqrt{z_{0}^{2}-z^{2}}}\nonumber\\
&=&-{4\over 5-p}J\left\{ \log {1\over z_{0}\Lambda}
-\log (1+\sqrt{1-{1\over z_{0}\Lambda}})\right\}.
\end{eqnarray}
In the limit $1/\Lambda\ll z_{0}$, the first term dominates, 
and we obtain
\begin{equation}
\bar{S}\sim {4\over 5-p}J\log (z_{0}\Lambda).
\end{equation}
Using (\ref{eq:tfti}), 
the classical contribution to the amplitude becomes
\EQ
\langle t_f,1/\Lambda,J ;T_{b}| t_i,1/\Lambda ,J ;-T_{b}\rangle_{class} =e^{-\bar{S}}
=\left( 2 \over \Lambda |t_{f}-t_{i}|\right)^{{4\over 5-p}J}.
\EN
The power law behavior with respect to 
the target (coordinate) 
distance $|t_{f}-t_{i}|$ conforms to the predictions of 
generalized conformal symmetry, indicating that the leading 
behavior of 
$\Delta_I$'s  for large $J \, (\sim q_p^{1/2})$ are $4J/7$,
for the case of the D0-branes. 
This is consistent with the supergravity result mentioned 
in the Introduction, and already suggests 
that the theories at the boundary corresponds to  nontrivial 
infra-red fixed points. Note also that in spite of the 
difference from the D3-case with respect to the 
affine time interval, the behavior of two-point 
functions with respect to the cutoff parameter shows 
similar power-law structure. 
In view of this power-law behavior, 
we can think of $\ln z \rightarrow \ln(1/\Lambda)$, 
rather than the original affine time,  as the 
effective time parameter along the 
tunneling trajectory near the boundary.

\subsection{Fluctuations around the tunneling null geodesics}

We now derive the action for fluctuations around 
the tunneling null geodesic. 
In the present paper, we study only the 
bosonic part, postponing the treatment 
of fermionic excitations and supersymmetry 
to a forthcoming work. 
It is convenient to use the following form of the space-time 
metric
%(doublly Wick rotated, $(\tau, \psi)
%\rightarrow -i(\tau, \psi)$):
\EQA
ds^{2}&=&q_p^{1/2}\Big[-dv\left(2du-H^{-1/2}dv+2\ell H^{-1/2}dx\right)
+(\ell^{2}H^{-1/2}-H^{1/2}r^{2})dx^{2}\nonumber\\
&&\quad
+H^{-1/2}d\tilde{x}_{a}^{2}+H^{1/2}r^{2}\cosh^{2}
\psi d\Omega_{7-p}^{2}\Big]
.
\label{eq:uvzmetric}
\EQN
The metric (\ref{Dpmetric}) (doubly Wick rotated
$(t, \psi)\rightarrow -i(t, \psi)$) 
can be brought to the above form \cite{blauetal} 
by the coordinate transformations, 
\[
u=u(r),\qquad
v=t+\ell\psi+a(r),\qquad
x=\psi+b(r),
\]
with
\[
{du\over dr}=\pm {H^{1/2}\over \sqrt{{\ell^{2}\over r^{2}}-H}},\qquad
{da\over dr}=\mp \sqrt{{\ell^{2}\over r^{2}}-H},\qquad
{db\over dr}=\mp{\ell\over r^{2}}{1\over \sqrt{{\ell^{2}\over r^{2}}-H}}
.
\]
%Using this metric and substituting the solution to the 
%standard string (bosonic) action 
Substituting this metric into the 
standard string (bosonic) action 
\begin{equation}
S={1\over 4\pi}\int d\tau \int_{0}^{2\pi\alpha} d\sigma
(\partial_{\tau}x^{\mu}\partial_{\tau}x^{\nu}g_{\mu\nu}
+\partial_{\sigma}x^{\mu}\partial_{\sigma}x^{\nu}g_{\mu\nu}
+i\epsilon^{\alpha\beta}\partial_{\tau}x^{\mu}\partial_{\sigma}x^{\nu}
B_{\mu\nu}), 
\label{eq:staction}
\end{equation}
it is easy to check that the trajectory defined by 
\begin{equation}
u= u^{(0)}\equiv \tau ,\quad v=v^{(0)}\equiv 0,\quad x=x^{(0)}\equiv 0,
\quad \tilde{x}_{a}=\tilde{x}_{a}^{(0)}\equiv 0, 
\end{equation}
satisfies the equations of motion and the Virasoro constraint 
\EQ
-\partial_{\tau}x^{\mu}\partial_{\tau}x^{\nu}g_{\mu\nu}
+\partial_{\sigma}x^{\mu}\partial_{\sigma}x^{\nu}g_{\mu\nu}=0, 
\label{eq:constr1}
\EN
\EQ
\partial_{\tau}x^{\mu}\partial_{\sigma}x^{\nu}g_{\mu\nu}=0.
\label{eq:constr2} 
\EN
We choose the string length parameter 
$\alpha \sim P^+$ 
to be proportional to the target-space energy,  $\alpha=E/q^{1/2}_{p}=J/(\ell q^{1/2}_p)$, 
such that the definitions of angular momentum $J$ and 
the target energy  $E$ coincide with the 
convention as introduced in subsection {\it 2.1}.

The fluctuations around the trajectory are 
treated by performing expansion 
\[
x_{\mu}=x^{(0)}_{\mu}+L x^{(1)}_{\mu}+L^{2} x^{(2)}_{\mu}+\cdots.
\]
Then, the
${\cal O}(L^{0})$ part (classical part)
of the action 
vanishes due to the constraints\footnote{For the present 
purpose, we consider the Routh function, 
which gives non-zero contribution at ${\cal O}(L^{0})$,
as we have seen.
Since the term $-J\int d\tau\partial_{\tau}(\psi_{0}+L\psi_{(1)}+\cdots)$
added to the original action in order to convert to the Routh function 
is a total derivative, it does not affect the bulk 
action for the fluctuations.}
and ${\cal O}(L^{1})$ part vanishes since $x^{(0)}$
satisfies the equation of motion. 
To the next order, the Virasoro constraints 
require 
\[
v^{(1)}{}'=\dot{v}^{(1)}=0, 
\]
and the action reduces to 
\begin{eqnarray*}
S^{(2)}&=&{1\over 4\pi}
\int d\tau \int_{0}^{2\pi E/q_p^{1/2}} d\sigma
\Big\{ H^{-1/2}(\dot{\tilde{x}}_{a}{}^{2}+\tilde{x}_{a}'{}^{2})
%\partial_{\alpha}x_{a}^{(1)}\partial^{\alpha}x_{a}^{(1)} 
+(\ell^{2}H^{-1/2}-H^{1/2}r^{2})
(\dot{x}{}^{2}+x'{}^{2})\nonumber\\
%\partial_{\alpha}z^{(1)}\partial^{\alpha}z^{(1)}\nonumber\\
&&\hspace{3cm}+H^{1/2}r^{2}\cosh^{2}b
(\dot{y}_{l}{}^{2}+y_{l}'{}^{2})\Big\}
%\partial_{\alpha}y_{l}^{(1)}\partial^{\alpha}y_{l}^{(1)}\Big\}
,
\end{eqnarray*}
where we have suppressed the superscript $(1)$ on the fields, and
$y_{l}$ $(l=1,\ldots, 7-p)$ are the coordinates along  the  
sphere ($S^{7-p}$) directions.  
We set the expansion parameter 
as $L=q_p^{-1/4}$. The higher order terms 
are neglected in the limit of  large $q_p$. 
Here, $H$ and $r$ are evaluated on the classical trajectory 
and hence depend on $\tau$.
Performing the field redefinition
\[
\tilde{x}_{a}\to H^{1/4}\tilde{x}_{a}, \qquad
y_{l}\to {1\over H^{1/4}r \cosh b}y_{l},\qquad
x\to {1\over \sqrt{\ell^{2}H^{-1/2}-H^{1/2}r^{2}}}x,
\]
and partial integration, the action becomes 
\begin{equation}
S^{(2)}={1\over 4\pi}\int d\tau \int_{0}^{2\pi \alpha}
\hspace{-0.5cm}d\sigma
\Big\{ \dot{\tilde{x}}_{a}{}^{2}+\tilde{x}_{a}'{}^{2}
+m^{2}_{\tilde{x}}(\tau)\tilde{x}_{a}^{2}
+\dot{x}{}^{2}+x'{}^{2}
+m^{2}_{x}(\tau)x^{2}
+\dot{y}_{l}{}^{2}+y_{l}'{}^{2}
+m^{2}_{y}(\tau)y_{l}^{2} \Big\},
\label{eq:Dpfluc}
\end{equation}
where 
\begin{eqnarray}
m_{\tilde{x}}^{2}&=&m_{x}^{2}=
-{(7-p)\over 16r^{2}}
\left\{(3-p)+(3p-13)\ell^{2}r^{5-p}\right\},\\
m_{y}^{2}&=&
-{(7-p)\over 16r^{2}}
\left\{(3-p)-(p+1)\ell^{2}r^{5-p}\right\}.
\end{eqnarray}
These formula for the mass parameters
 have been previously obtained \cite{gimonetal} 
in the real affine-time approach. 
Here $r=r(\tau)$ is given by the 
classical solutions determined by (\ref{eq:drdtau}). 
Hence the masses are time dependent except for $p=3$. 
In particular, for $p<3$, the masses asymptotically 
increase indefinitely, while for $5>p>3$ they vanish 
asymptotically, suggesting physical consequences 
drastically different  from the conformal case for 
both cases. 
Intuitively, the former corresponds to the fact that the 
external tidal force defeats the string tension 
and the latter to the converse that the 
effect of the tidal force becomes negligible,  
as the strings approach the boundary.  

There remain no  fields with negative metric 
in the Euclidean sense. 
On the other hand, 
the squared masses can become negative in the region 
deep inside the bulk. Near the boundary 
they are always positive for $0\le p \le 4$ and asymptotically 
become infinite. 
By rewriting the above expressions using 
turning point $r_0$,  
\[
m^{2}_{\tilde{x}}=m^{2}_{x}={(7-p)\ell^2\over 16r^{2}}[2(5-p)r_{0}^{5-p}
+(13-3p)(r^{5-p}-r_{0}^{5-p})],
\]
\[
m^{2}_{y}={(7-p)\ell^2\over 16r^{2}}[2(p-1)r^{5-p}_{0}+(p+1)(r^{5-p}-r^{5-p}_{0})],
\]
 we see that $m_{y}^{2}$ is positive for $1\le p \le 4$
in the allowed region. 
 It should be emphasized here that if we remain 
in the usual picture of real null geodesics 
we would have been confronted by the 
difficulty that  (mass)$^2$ are negatively infinite 
as we approach the horizon.

Note 
that the action (\ref{eq:Dpfluc}) has a global 
$SO(p+1)\times SO(7-p)$ symmetry 
and also more importantly
 that the action is invariant under the 
scaling transformation 
\EQ
r\rightarrow \lambda r, \quad \ell\rightarrow \lambda^{-(5-p)/2}\ell, \quad (\tau, \sigma) 
\rightarrow (\lambda\tau, \lambda\sigma), \quad 
\alpha \rightarrow \lambda\alpha
\label{scaling}
\EN 
with all the fluctuating fields being of zero dimension. 
This property of course reflects  the pseudo-symmetry under 
the generalized scaling transformation of the 
original metric, which for general $p$ takes the form 
$X_i \rightarrow \lambda X_i, \quad \vec{x} \rightarrow \lambda^{-1}\vec{x}, \quad g_s \rightarrow \lambda^{3-p}g_s$, where 
$\vec{x} \propto (t, \tilde{x}_a)$ 
and $X_i$ are base space and 
transverse coordinates, respectively. The scaling of 
$\ell \, (\sim |\vec{x}_1-\vec{x}_2|\equiv 
|t_f-t_i|)$ is the consequence of our rescaling of the 
base space coordinate $\vec{x}\rightarrow q_p^{1/2}\vec{x}$. 
As in the ordinary conformal case, 
the cutoff $\Lambda$ in general breaks this scaling 
property. However,  after the dependence on the cutoff 
is suitably eliminated by wave function renormalization, the correlation functions 
should be symmetric under these scaling transformations. 
This is indeed satisfied 
by the results of supergravity analyses \cite{sekiyone} for $p=0$. The symmetry under the scaling transformation (\ref{scaling}) will play an important role in the present work too. 

Finally, we note that, for the purpose of the present paper 
discussing the two-point transition amplitudes 
within the string-tree approximation, the dilaton 
background $\phi$ can be ignored. The reason is that 
 its coupling to 
strings occurs only through the world-sheet 
curvature term $\int d\tau d\sigma \sqrt{h^{(2)}}\phi R^{(2)}
$ which in the limit $L\rightarrow 0$ reduces 
to  $\int d\tau d\sigma \sqrt{h^{(2)}}\phi_{classical}\\
\times R^{(2)}$ and hence 
does not couple with the fluctuating 
fields.  But it would play an important role 
for ensuring the world-sheet conformal symmetry 
by cancelling the anomaly, especially  
when we discuss the string-loop effects. 
Note also that,  
provided the world-sheet conformal symmetry 
is valid, the world-sheet metric can be chosen 
such that $R^{(2)}=0$ for the world sheets of 
cylinder topology.

\section{General theory of harmonic oscillators with 
time dependent masses}
\setcounter{equation}{0}
In the conformal case of $p=3$,  
the quantization of our system \cite{dobashimayone} 
is completely 
straightforward since the mass is constant, and the 2-pt 
amplitude takes the trivial form 
\[
\langle t_f,1/\Lambda,J,\{N_{n}\} ;T| t_i,1/\Lambda,J,\{N_{n}\} ;-T\rangle
=\left({1\over \Lambda |t_{f}-t_{i}|}\right)^{2 (J+\overline{\Delta})},
\]
where, as in the usual formulation \cite{bmn} using a 
real null geodesic,  $\overline{\Delta}=\sum_{n}N_{n}\sqrt{1+{n^{2}R^{4}\over J^{2}}}$ is the quantum contribution to 
conformal dimension, corresponding to the 
frequency $\omega_{n}= \sqrt{\ell^{2}+{R^{4}n^{2}\over E^{2}}}$ of the modes of strings, each of  which 
contributes to the amplitude as $e^{-2\omega_{n}T}$.  
Here we have subtracted the zero-point energies, since 
they cancel after taking into account the fermionic 
excitations. 
In the nonconformal cases $p\ne 3$, 
the situation is much more 
nontrivial, owing to the complicated time 
dependence of mass functions. 
Thus, to study the fluctuations around the classical 
trajectory, we now have to 
develop a general quantum 
theory of harmonic oscillators with time-dependent 
potential of the form $m(\tau)^2x(\tau)^2/2$. 
We hope that our discussion will be useful for 
other cases, such as cosmological 
applications \cite{paparussotsey}, 
with time dependent backgrounds than the 
present specific example. Hasty readers may wish to 
skip general formalism below and go directly to the 
final formula (\ref{finalsope}), which is astonishingly 
simple. 

We start from  considering the quantization 
of coordinate operator satisfying the ({\it Euclidean}) 
equation of motion in the Heisenberg representation, 
\EQ
{d^2 \over d\tau^2}X(\tau)=m(\tau)^2X(\tau)
. 
\EN
In our cases, by choosing the origin of time appropriately, we can assume $m(\tau)=m(-\tau)$.  
The general solution can be expressed in the form
\EQ
X(\tau)=f_+(\tau)a + f_-(\tau)a^{\dagger}
\EN
with the normalization condition 
for the Wronskian 
\EQ
f_+{df_-\over d\tau}-f_-{df_+\over d\tau}=1, 
\label{wronskian}
\EN
which is possible since 
\EQ
{d\over d\tau}(f_+{df_-\over d\tau}-f_-{df_+\over d\tau})=0. 
\EN
The solutions $f_{\pm}(\tau)$ are chosen such that 
they satisfy the boundary condition 
\EQ
f_{\pm}(\tau)\rightarrow 0, \quad \tau \rightarrow 
\pm T_b ,
\EN
near the boundary $r\rightarrow \infty$ in the 
large $q_p$ limit.  They corresponds to positive and 
negative frequency solutions,  respectively, 
when the Euclidean affine 
time $\tau$ is formally Wick-rotated to the 
real affine time $(\tau \rightarrow it)$ in the case 
of constant mass. We stress that 
naive approximate methods for obtaining $f_{\pm}(\tau)$, 
such as adiabatic or WKB-like treatments,  
are not allowed, since  
we cannot assume that $d m(\tau)/d\tau \ll 
m(\tau)^2$: For large $r$, $dm(\tau)/d\tau$ and $m(\tau)^2$ 
are in general of the same order $O(r^{3-p})$ $(p\ne 3)$.  

Note that $a, a^{\dagger}$ are assumed to be 
independent of $\tau$. 
By using the time reflection symmetry, we can set 
\EQ
f_-(\tau)=f_+(-\tau)
,
\EN
which means that $X$ satisfies the 
reflection condition
\EQ
X(\tau)^{\dagger}=X(-\tau)
,
\label{reflcond}
\EN
instead of the ordinary condition of hermiticity in the 
real-time formulation. Here we have assumed that 
the solutions $f_{\pm}$ are real. When $m(\tau)^2<0$ for 
some regions of time as in our cases, it could be that 
we were forced to use complex solutions. 
In such a situation, the time-reflection 
condition should be replaced by 
\EQ
f_-(\tau)=\overline{f_+(-\tau)}, 
\EN
keeping the reflection condition (\ref{reflcond}). 
Actually, it will turn out that we can  assume 
real solutions for our later 
applications. 

%\noindent
The quantization condition is  expressed as 
\EQ
[X(\tau), P(\tau)]=i
,
\EN
by defining the momentum operator
\EQ
P(\tau)=i{d\over d\tau}X(\tau)=i({df_+\over d\tau}a 
+{df_-\over d\tau}a^{\dagger} )
, 
\EN
 which  satisfies the same reflection 
condition as the coordinate operator,  
$
P(\tau)^{\dagger}=P(-\tau)
.
$
The quantization condition is equivalent to 
\EQ
[a, a^{\dagger}]=1
,  
\EN
because of the Wronskian condition (\ref{wronskian}). 

Let us now try to transform this formalism into 
the Hamiltonian picture. 
First, we can rewrite 
the equation of motion in the first order form as 
\EQ
i{dX\over d\tau }=P, \quad i{dP\over d\tau}=-m(\tau)^2X
\EN
which is expressed using commutator
as
\EQ
{dX\over d\tau}=[H(\tau), X], \quad {dP\over d\tau}
=[H(\tau), P]
\EN
with the time-dependent Hamiltonian, 
\EQ
H(\tau)={1\over 2}(P^2 + m(\tau)^2X^2)
. 
\EN
We can then introduce the transition operators by, 
depending respectively on positive and negative $\tau$,\EQ
U_+(\tau,0)={\cal T}_+\exp[\int^{\tau}_0 d\tau' H(\tau')]
\EN
or 
\EQ
U_-(-\tau,0)={\cal T}_-\exp[-\int^{\tau}_0 d\tau' H(-\tau')]
\EN
for $\tau>0$,  with ${\cal T}_{\pm}$ being the time or 
anti-time ordering operation, respectively. They satisfy 
\EQ
X(\pm\tau)=U_{\pm}(\pm\tau, 0)X(0)U_{\pm}(\pm\tau,0)^{-1}, \quad 
P(\pm\tau)=U_{\pm}(\pm\tau,0)P(0)
U_{\pm}(\pm\tau,0)^{-1}
, 
\EN
for $\tau\ge 0$. 
The reflection condition for the transition operator 
takes the form 
\EQ
U_{\pm}(\pm\tau, 0)^{\dagger}=U_{\pm}(0, \mp\tau)
.
\EN
By definition, we have also 
\EQ
U_-(-\tau, 0)^{-1}=U_+(0, -\tau)
 .
\EN

In the case of ordinary time-independent harmonic 
oscillator, we have $U_{\pm}(\pm\tau, 0)=\exp(\pm\tau H)=
U_{\pm}(0, \mp\tau)$; namely $U_{\pm}$ are 
diagonalized and hermitian.  
In the general time-dependent cases,  
however, the transition operators 
are neither diagonalized, nor hermitian in the usual sense.    

The relation 
between Heisenberg and Schr\"{o}dinger pictures  
is formulated as
\EQ
\langle \psi_1| O(\tau)| \psi_2\rangle=
\langle \psi_1(\tau)| O| \psi_2(\tau)\rangle. 
\EN
Then the Euclidean {\it out} and {\it in} (ket) states 
in the Schr\"{o}dinger picture are defined  using the 
above transition operators as 
\EQ
|\psi_f(\tau)\rangle_{out}=U_+(\tau, 0)^{-1}|\psi_f\rangle
,
\EN
\EQ
|\psi_i(-\tau)\rangle_{in}=U_-(-\tau, 0 )^{-1} |\psi_i\rangle
,
\EN
for $\tau= T\rightarrow T_b$. 
Note that the bra-states $\langle \psi(\pm\tau)|$ 
are in general 
{\it not} the conjugate of the ket-states 
$|\psi(\pm\tau)\rangle$ 
 in the above relation, 
since the former are defined as 
\[
\langle \psi(\tau)|=\langle \psi|U_+(\tau,0), 
\quad 
\langle \psi(-\tau)|=\langle \psi|U_-(-\tau,0)
.
\]
Note however that 
these bra-and-ket states are 
defined at the common time 
$\tau=0$. 
The orthonormality 
condition can then be expressed 
in terms of this internal product, which is by definition 
independent of $\tau$, 
as 
\[
\langle \psi_1|\psi_2\rangle =\langle \psi_1(\pm\tau)|\psi_2(\pm\tau)\rangle =\delta_{12}. 
\]
The initial and final states,  
$|\psi_i\rangle$ and $|\psi_f\rangle$, are 
defined at times, $\tau=-T$ and $\tau=T$, 
respectively. 
In this circumstance, 
the  natural definition of the S-matrix is 
\[
S_{fi}\equiv\, \,  
_{in}\langle\psi_i(-\tau)|\psi_f(\tau)\rangle_{out}
=\langle\psi_i|U_-(-\tau,0)
U_+(\tau, 0)^{-1}|\psi_f\rangle
\]
\EQ
 =\langle\psi_i|\Big(U_+(\tau,0)U_-(-\tau, 0)^{-1}\Big)^{-1}|\psi_f\rangle=
\langle\psi_i|U_+(\tau,-\tau)^{-1}|\psi_f\rangle
\EN
with
\EQ
U_+(T, -T)={\cal T}_+\exp[\int_{-T}^{T} d\tau
H(\tau)]\equiv S^{-1}(T)
.
\EN
Thus, 
\EQ
S(T)={\cal T}_-\exp[-\int_{-T}^{T}d\tau
H(\tau)]
.
\EN
Due to the time-reflection condition, the 
$S$-operator is  hermitian, in contrast to 
general transition operators $U$'s. 
For time independent case, 
this leads to the naive Euclideanized S-operator $S=\exp (-2TH)$. 

The main task for obtaining the two-point 
functions is then to `diagonalize' the S-matrix operator, 
$S(T)$.   Our strategy toward this goal is as follows: 
Since the time dependent Hamiltonian is 
quadratic  in the time-independent 
$(a, a^{\dagger})$ basis, we can always express the 
S-operator in the following normal-ordered form:
\EQ
S(T)=N(T):\exp[{1\over 2}A(T)(a^{\dagger})^2
+B(T)a^{\dagger}a + {1\over 2}C(T)a^2]:
~.
\label{normalform}
\EN
Because of hermiticity, $A=C$  and $B$ is real. 
This can further be converted to the 
exponential form 
\EQ
S(T)=\tilde{N}(T)\exp[{1\over 2}\tilde{A}(T)(a^{\dagger})^2
+\tilde{B}(T)a^{\dagger}a +{1\over 2}\tilde{C}(T)a^2]
,
\label{expoform}
\EN
which can then be transformed to a `diagonalized' form 
\EQ
S(T) \rightarrow 
{\cal N}(T)\exp[-\Omega(T)b^{\dagger}(T)b(T)]
\EN
with 
\EQ
[b(T), b^{\dagger}(T)]=1
\EN
by a suitable 
$T$-dependent Bogoliubov 
transformation $(DG-EF=1)$, 
\EQ
(a, a^{\dagger}) 
\rightarrow (b(T)=D(T)a + E(T)a^{\dagger}, b^{\dagger}(T)=
F(T)a+G(T)a^{\dagger})
. 
\EN
This immediately leads to 
an expression for $\Omega(T)$ and the normalization, 
\EQ
\Omega=\sqrt{\tilde{B}^2-\tilde{A}\tilde{C}}
,
\EN
\EQ
{\cal N}=\tilde{N}\exp[-{\tilde{B}(T)+\Omega(T) \over 2}]
.
\EN
Thus, the S-operator can always be 
diagonalized for {\it arbitrary} $T$ when the 
initial ket-states $|\psi_i\rangle$ and 
the final bra-states $\langle \psi_f|$ 
are represented in the Fock bases 
$\{b^{\dagger}(T)^n|0\rangle_b\}$ and 
$\{\, _b\langle 0|b(T)^n\}$, 
respectively, where the `vacuum' states are 
defined by  $b(T)|0\rangle_b=0
=\, _b\langle 0|b^{\dagger}(T)$~.
In terms of the original creation and 
annihilation operators, these vacua are 
`squeezed' states such as
$|0\rangle_b \sim
\exp (-{E(T)\over D(T)}(a^{\dagger})^2)|0\rangle $. 
Because of time-dependent potential, quanta corresponding 
to $a, a^{\dagger}$ are being 
continuously created and annihilated during propagation 
along the trajectory. But, the transition amplitudes  can be 
diagonalized in suitable bases prepared at the 
both initial and final states constructed as above. 
To avoid possible confusion, we emphasize that 
what we are doing here is entirely different from the 
diagonalization of the Hamiltonian itself at 
each given time, which does not provide 
any help for our purpose. Our procedure 
takes the dynamical transition of states from $\tau=-T$ to $\tau=T$ into full account.  
This kind of `integrated' diagonalization 
has not been familiar 
in the preceding literature. However, 
this  is  the analogue in the first-quantized 
approach to  the 
diagonalization of linearized fluctuations in 
arbitrary backgrounds in (second-quantized) 
field theory formalism. 

 Here we collect some of key formulae for performing 
these manipulations. 
In terms of the solutions $f_{\pm}$, the 
coefficient functions of the normal ordered form 
are given by solving the following differential equations 
with the boundary conditions 
$N(0)=1, A(0)=B(0)=0$:
\EQ
{d\over dT}\ln(1+B)={d\over dT}\ln N^2=-2\beta-2\alpha A
,
\EN
\EQ
{d\over dT}A=-(\gamma+2\beta A +\alpha A^2 )
-\alpha (1+B)^2
,
\EN
with
\EQ
\alpha=m^2 f_-^2-\dot{f}_-^2, 
\quad
\beta=m^2f_+f_- - \dot{f}_+\dot{f}_-, \quad
\gamma=m^2f_+^2-\dot{f}_+^2, 
\EN
which follow from the
equation 
\EQ
{dS(T)\over dT}=
-H^{\dagger}(T)S(T)-S(T)H(T) 
.
\EN
We have assumed that the solutions $f_{\pm}$ are 
all real. 

In fact, apart from the normalization function $N(T)$, 
the transition operator 
can be algebraically determined 
by the following simple trick. First note that by its definition 
the $S$ operator satisfies the commutation relations with the coordinate and 
momentum operators, 
\EQ
X(T)=S(T)^{-1}X(-T)S(T), \quad 
P(T)=S(T)^{-1}P(-T)S(T), 
\EN
which, using the normal order form of the operator 
$S(T)$ with $A=C$,  reduce   to  
\EQ
1+B-A={f_+(T)\over f_-(T)}, \quad 
1+B+A=-{\dot{f}_+(T)\over \dot{f}_-(T)}
\label{ba}
, 
\EN
respectively. 
It is easy to check that these forms indeed satisfy 
the above differential equations with the initial 
condition $A(0)=B(0)=0$. 
Then, combining these formulae 
with the result for the normalization function 
obtained from the differential equation, 
we  have the completely explicit form for the 
normal-ordered $S$-operator  as 
\EQ
 N^2=1+B={1\over 2f_-(T)\dot{f}_-(T)}, 
\EN
\EQ
A=-{1\over 2}\Big(
{f_+(T)\over f_-(T)} +{\dot{f}_+(T)\over \dot{f}_-(T)}
\Big) 
\label{at}
.
\EN

The coefficient functions of the exponential form 
are then determined by the following 
general formulae relating 
normal and exponential forms, (\ref{normalform}) and 
(\ref{expoform});
\EQ
\exp\pmatrix{-\tilde{B} & -\tilde{A} \cr 
                           \tilde{C} & \tilde{B}\cr}
=\pmatrix{{1\over 1+B} & {-A\over 1+B} \cr
                        {C\over 1+B} & 1+B -{CA\over 1+B}\cr}, 
\EN
\EQ
\tilde{N}(T)=N(T)\sqrt{1+B}\exp(-\tilde{B}/2)
=(1+B)\exp(-\tilde{B}/2). 
\EN
For example, the former equation is derived by computing 
$$
\pmatrix{ S a S^{-1}\cr  S a^{\dagger} S^{-1}}
={\cal L} \pmatrix{ a \cr a^{\dagger}}
$$
using the two (normal and exponentiated) 
expressions of $S$ and equating the results, 
where ${\cal L}$ is the $2\times 2$ matrix defining 
the Bogoliubov transformation corresponding to 
the S-operator. In particular, the eigenvalue 
function $\Omega$ for the diagonalized S-matrix is given by
\EQ
\cosh\Omega =\frac{1}{2}
\left(
1+B+\frac{1-AC}{1+B}
\right).
\EN
All these formula are exact for arbitrary $T$, and, 
if we wish,  
can trivially be converted to real affine-time 
formulation.  
In the special case of constant mass, we have the familiar formulas as 
\[
\alpha=\gamma=0, \quad \beta=m\quad \rightarrow 
\quad A=C=0=\tilde{C}=\tilde{A}, \quad \Omega=-\tilde{B}
, \quad \e^{\tilde{B}}=1+B=e^{-2mT}, 
\]
and hence $S=\e^{-m(2a^{\dagger}a+1)T}$.

Finally, we can  see the crucial role played by the 
boundary conditions for $f_{\pm}$. Since 
$f_+(T)=f_-(-T)\rightarrow 0$ in the limit of large 
$r(T)$ ($T\rightarrow T_b$), 
(\ref{ba}) shows 
that $A(T) \sim1+B(T) \rightarrow 0$ asymptotically 
near the 
boundary. This 
leads to the simple exponential form of the 
boundary-to-boundary S-operator, 
\EQ
S(T)\rightarrow (1+B)^{a^{\dagger}a +1/2}
=(2f_-(T)\dot{f}_-(T))^{-(a^{\dagger}a +1/2)}. 
\label{finalsope} 
\EN 
Namely, the time-independent creation 
and annihilation operators $(a^{\dagger},a)$ themselves 
diagonalize the 2-pt functions 
as in the case of constant mass; 
$(b^{\dagger}(T), b(T))
\rightarrow (a^{\dagger}, a) $, as we approach the 
boundary $T\rightarrow T_b$. 
This is remarkable in view of 
the complicated time dependence, but is 
owing to our choice  
of  representations for the 
coordinate and the momentum 
with our boundary condition for $f_+$ together 
with the time-reflection symmetry.  
For this remarkably simple result, the Euclidean nature 
of our system is very important. If we consider the case 
of real affine time, the coefficient function $A$ 
provides an oscillating 
contribution whose frequency is of the same order 
as that of $1+B$, and hence it is not clear whether 
the corrections coming from $A$ can safely be 
neglected.

\section{Two-point PP-wave S-matrix for D$p$-brane backgrounds}
\setcounter{equation}{0}

We now proceed to the derivation of diagonalized two-point 
S-matrices for  D$p$-brane backgrounds on the basis 
of general formalism given in the previous section.  
We first consider the case of D0-brane  within 
supergravity approximation ignoring stringy excitations and 
confirm that the results are consistent with the previous 
field-theory analysis. 
The case of general $p \,  (<5)$ including the stringy modes will then 
be treated. 
\subsection{Case of D$0$ in supergravity approximation without stringy excitations}
 The equation of motion we have to solve is 
\EQ
{d^2 \over d\tau^2}X(\tau)=m(\tau)^2X(\tau)
,
\label{eq:fieldeq}
\EN
with mass function 
 \EQ
m(\tau)^{2}={7\over 16}{(\ell^{2}r^{5}-3 ) 
\over  r^{2}}\,(=m_y^2 )
,
\EN
or 
\EQ
m(\tau)^{2}={7\over 16}{(13\ell^{2}r^{5}-3  ) 
\over r^{2}}(=m_x^2), 
\EN
and radial function $r=r(\tau)$ satisfying 
\EQ
\frac{dr}{d\tau} =
\left\{
\begin{array}{@{\,}ll}
\sqrt{ \ell^2 r^{5}-1} & (\tau>0),
\\
-\sqrt{ \ell^2 r^{5}-1} & (\tau<0).
\end{array}
\right.
\EN
The turning point is $r_0=\ell^{-2/5}$.  Since $\tau$ is 
only given as an implicit function, 
it is more convenient to 
convert the equation of motion in terms of the radial 
coordinate $r$. We obtain 
\EQ
(\ell^2 r^5-1)  \frac{d^2 X}{dr^2} + \frac{5}{2} \ell^2 r^4   \frac{dX}{dr}  -m(r)^2 X=0 .
\EN
For the  mass functions above, this can be solved as 
\EQ
X_y(r) = C_1 \, r^{-3/4}
\exp \left(\ell 
 \! \int \frac {r^{3/2}}{\sqrt{\ell^{2}\,r^{5} - 1}}\,dr \! 
 \right) 
+ 
C_2 \, {r^{-3/4}}
\exp \left(  \!  - \ell  \int \frac {r^{3/2}}{\sqrt{\ell^{2}\,r^{5} - 1}}\,
dr \!  \right)  
\EN 
and 
\EQ
X_x(r)=C_1 \,r^{7/4} + C_2 \, \frac{\sqrt{\ell^{2}\,r^{5} - 1}}{r^{3/4}}
, 
\EN
respectively. 

We have to choose the basis of  the solutions satisfying 
our boundary conditions required in the previous section. 
Let us first 
take the second case with the mass function $m_x$.  
As a function of $\tau$, the solution must be 
continuous at $\tau=0$.  To obtain the 
solution conforming  to this requirement, 
we  can set
$
X_x(\tau)=\tilde{C_1} \,r^{7/4} + \tilde{C_2} \,  \frac{dr}{d\tau} {r^{-3/4}} ,
$
and determine the coefficients such that 
the boundary condition is satisfied. We find
\EQA
f_+^x(\tau) &=&  \sqrt{{\ell \over 5}}\left(
r^{7/4} - \frac{1 }{\ell} \frac{dr}{d\tau} r^{-3/4} \right)
,
\\
f_-^x(\tau) &=&  \sqrt{{\ell \over 5}}\left(
r^{7/4} + \frac{1 }{\ell} \frac{dr}{d\tau} r^{-3/4} \right)
.
\EQN
Here and in what follows, we adopt the same normalization 
condition as in the previous section. 
Similarly, for the case of mass function $m_y$, we find 
\EQA
f_+^y(\tau) &=&  {1\over \sqrt{2\ell}} \, {r^{-3/4}}
\exp \left(  \!  - \ell \int_0^\tau r(\tau)^{3/2}\, d\tau \!  \right)  
, 
\\
f_-^y(\tau) &=&    {1\over \sqrt{2\ell}}\, {r^{-3/4}}
\exp \left(   \ell \int_0^\tau r(\tau)^{3/2}\, d\tau \!  \right)  
.
\EQN
Here the integral on the exponential can be 
performed, giving (for $\tau>0$) 
\EQA
\ell \int_{0}^{\tau}  \, r(\tau)^{3/2} d\tau &=&
\int_{r_0}^{r} \frac{\ell\,r^{\frac {3}{2}}}
{\sqrt{\ell^{2}\,r^{5} - 1}}
\,dr
=
%\frac {1}{5} 
%\,\ln\left[
%2\,\sqrt{\ell^{2}\,r^{5}\,(\ell^{2}\,r^{5} - 1)}+ 2\,\ell^{2}\,r^{5} %- 1
%\right]
\frac {2}{5} 
\ln\left[
\sqrt{\ell^{2}\,r^{5}}+ \sqrt{\ell^{2}\,r^{5} \!-\! 1}\,
\right]
\nonumber
\\
&\rightarrow& \ln( \ell^{2/5}\,r) \quad (r \rightarrow \infty)
.
\EQN
We remark that, in spite of the fact that the masses squared  
become negative near the turning point, the exact solutions 
remain as real functions in the whole allowed range of the 
radial coordinate $r$.  

Given these solutions, we can immediately apply the 
general formulae of the previous section. 
The near-boundary behaviors of the coefficient functions 
$A, B$ are  
\EQA
A &\sim& \frac{3}{28 \ell^2} r^{-5} + {\cal O}(\ell^{-4}r^{-10}) , 
\\
B +1\, (=N^2)&\sim& \frac{5}{14 \ell^2} r^{-5} + {\cal O}(\ell^{-4}r^{-10}) 
,
\EQN
and 
\EQA
A &\sim& 3 \,(4\ell^2\, r^5)^{-2/5} + {\cal O}((\ell^{2}r^{5})^{-7/5}), 
\\
B +1\, (=N^2)&\sim& 4 \, (4\ell^2\, r^5)^{-2/5}
+
{\cal O}((\ell^{2}r^{5})^{-7/5})
,
\EQN
for the cases of $m_x$ and $m_y$, respectively.  
 Thus the contribution of each supergravity mode
to two-point functions  
 is, using the relation between $r(T)$ and the UV cutoff $\Lambda$, 
\[
{1\over \Lambda}={2\over 5-p}r(T)^{-(5-p)/2}, 
\]
\EQA
S_x(T) &\sim& (\ell^2\,r(T)^5)^{ - \left[a^{\dagger}a+ \frac{1}{2} \right]}   \, \, \,
\sim \, (|t_i-t_f|\Lambda)^{ -2\left[a^{\dagger}a
+ \frac{1}{2} \right]}
\label{so1scaling},
\\
S_y(T) &\sim& (\ell^2\,r(T)^5)^{ -\frac{2}{5} \left[a^{\dagger}a+\frac{1}{2} \right]} \,
\sim (|t_i-t_f|\Lambda)^{ -\frac{4}{5} \left[a^{\dagger}a+\frac{1}{2} \right]}
\label{so7scaling},
\EQN
depending on the $SO(1) $ or $SO(7)$ direction, respectively.

In order to make comparison with the  supergravity 
analysis given in \cite{sekiyone}, we have to 
multiply  these results over all modes of transverse 
directions. Therefore,
 fixing precisely the over-all constant term 
of the exponents 
 requires an accurate evaluation of 
zero-point contributions including the 
fermionic coordinates.  We leave such a full 
treatment of fermions and supersymmetry 
to a forthcoming work, and in the present work 
we will be satisfied by checking 
consistency with \cite{sekiyone},  
ignoring the constant part.  
Note that, once the dependence with respect to the 
target distance is fixed, the scaling symmetry guarantees that the two-point functions, with a prescribed normalization 
condition as discussed in the Introduction after 
removing the cutoff dependent factors,  have the 
correct behavior with respect to the coupling constant 
$q_0\sim g_sN$. 

Comparing the dependence on the 
target distance $|t_i-t_f|$ in the expressions 
(\ref{so1scaling}) and 
(\ref{so7scaling}) for the 
contributions to the S-operator with the 
general form of the two-point functions 
\EQ
\langle{\cal O}_I(t_1){\cal O}_I(t_2)\rangle
\sim {1\over g_s^2\ell_s^8}
(g_sN\ell_s^7)^{(\Delta_I+6)/5} |t_1-t_2|^{-(7\Delta_I+12)/5}, 
\EN
we see that a single excitation along the SO(1) or SO(7) 
direction contribute to the generalized conformal 
dimension as 
\EQ
\Delta_x= 10/7 \quad \mbox{or}\quad \Delta_y=4/7, 
\label{d0confdim}
\EN
respectively. 
On the other hand, as we have already mentioned 
 in the Introduction, the field-theory analysis predicts the 
spectrum for the generalized conformal dimension of the 
form
\EQ
\Delta_I =4\ell_I/7 + 2n_I-1
\EN
where  $\ell_I$ is the number of 
orbital excitations $\tilde{X}_i=X_i/q_0^{1/7} \, (i=1, 2, \ldots, 9)$ of transverse modes 
in the sense of M-theoretical 11 dimensions and 
$n_I=1-n_++n_-$ measures the numbers of 
11-dimensional (upper) light-cone indices. For example, 
$n_I$ and $\ell_I$ for 
the moments of the 11-dimensional energy-momentum tensor 
\EQAN
T^{++}_{i_1i_2\cdots i_k}&=&{\rm STr}(\tilde{X}_{i_1}\tilde
{X}_{i_2}\cdots\tilde{X}_{i_k}+\cdots)/g_s, \\
T^{+i}_{i_1i_2\cdots i_k}&=&{\rm STr}((D_0X_i)\tilde{X}_{i_1}\tilde
{X}_{i_2}\cdots\tilde{X}_{i_k}+\cdots)/g_s,\\ 
T^{+-}_{i_1i_2\cdots i_k}&=&{\rm STr}(
{1\over 2}(D_0X_i)(D_0X_i)\tilde{X}_{i_1}\tilde
{X}_{i_2}\cdots\tilde{X}_{i_k}+\cdots)/g_s,\\
&&\quad .... etc
\EQNN
are 
\[
(n_I, \ell_I)=(-1, k), \, (0, k), \, (1, k), \, .... etc
\]
For complete details on this correspondence, we refer 
the reader to  \cite{sekiyone}. For studying 
the BMN-type operators, 
we choose the direction of the angular momentum $J$ 
 to be $i=8, 9$ such that the BMN `$Z$' coordinate 
is 
\EQ
Z=(X_8+iX_9)/\sqrt{2}
\EN
and, correspondingly, the excitation modes 
along $SO(7)$ and $SO(1) $ are 
\EQ
\tilde{X}_i \, \, \mbox{with} \,\, i=1, 2, \ldots, 7 
\quad \mbox{or} \quad D_0Z , 
\EN
respectively: The BMN-ground state with a definite 
angular momentum $J$ 
corresponds to $T^{++}_{ZZ\cdots Z}$\, \,  $(J=\ell_I)$. 
 Then, starting from this ground state,  
single excitations along the $SO(7) $ and  $SO(1)$  
contribute to the shifts  $(n_I, \ell_I)\rightarrow 
(n_I, \ell_I+1)$ and  
$(n_I, \ell_I)\rightarrow (n_I+1 , \ell_I-1)$,    
respectively. Here the shift of $n_I$ in the second case 
originates from either $n_+\rightarrow n_+-1$ or $n_- \rightarrow n_-+1$, depending on operators 
in consideration. Hence,  the shift of the 
generalized conformal dimensions are 
$4/7$ and $2-4/7=10/7$, respectively, in agreement 
with the results (\ref{d0confdim}) of the particle-string picture in the PP-wave limit.

\subsection{General case of D$p \, \, (p<5)$ with stringy excitations}

Passing through the important nontrivial
 test that the PP-wave limit correctly 
reproduces the supergravity results for $p=0$ case, 
we are now in the position to study the general case 
of D$p$-brane backgrounds with stringy excitations. 
We treat each Fourier mode separately. Thus 
for the $n$-th stringy modes, we set 
$X(\tau, \sigma)\sim {1\over \sqrt{\pi\alpha}}\cos( \frac{n}{\alpha} \sigma) X(\tau)$ 
or ${1\over \sqrt{\pi\alpha}}\sin( \frac{n}{\alpha} \sigma) X(\tau)$, and the equation of motion takes the form, 
\EQ
\left(
{d^2 \over d\tau^2} -\frac{n^2}{\alpha^2}
\right)X(\tau)  =m(\tau)^2 X(\tau) 
\label{eq:eomstring}
\EN
with mass functions 
\EQ
m(\tau)^{2}=-
{(7-p) \over 16  r^{2}}
[ (3-p) +(3p-13)\ell^{2}r^{5-p}  ]
\, \, \, 
(=m_{p,x}^2)
,
\EN
or
\EQ
m(\tau)^{2}=-
{(7-p) \over 16  r^{2}}
[ (3-p) -(p+1)\ell^{2}r^{5-p}  ]
\,\, \, (=m_{p,y}^2 ) ,
\EN
where $r=r(\tau)$ is given by
\EQ
\frac{dr}{d\tau} =
\left\{
\begin{array}{@{\,}ll}
\sqrt{ \ell^2 r^{(5-p)}-1} & (\tau>0)
,
\\
-\sqrt{ \ell^2 r^{(5-p)}-1} & (\tau<0)
.
\end{array}
\right.
\EN
Note that in our convention 
the component field $X(\tau)$ has always 
scaling dimension $1/2$ under the 
scaling transformation (\ref{scaling}).  
This is necessary for the consisitency  with the 
normalization condition for $X(\tau)$ assumed 
in the previous section. 

For $n=0$, the equation of motion (\ref{eq:eomstring}) for each $m_{p,x}$
or $m_{p,y}$ is 
exactly solvable with the results for $\tau\ge 0$ 
as 
\begin{eqnarray}
f_{\pm}^x(\tau) &=& (5\!-\!p)^{-{1\over 2}} \ell^{-{1\over 5-p}}  u^{{p-3 \over 4(5-p)}}
( \sqrt{u} + \sqrt{u\!-\!1} )^{\mp 1} ,
\\
f_{\pm}^y(\tau) &=& 2^{-{1\over 2}} \ell^{-{1\over 5-p}} u^{{p-3 \over 4(5-p)}}
( \sqrt{u} + \sqrt{u\!-\!1} )^{\mp{2\over 5-p}}  ,
\end{eqnarray}
where $u=\ell^2 r^{5-p}$. 
These general expressions are valid 
for all $p<5$ including $p=0$ and $p=3$. 
The asymptotic behaviors of the above solutions for $r\rightarrow \infty$ ($\tau\rightarrow T_b$)
are given respectively as 
\begin{equation}
f_+^x \sim {1\over 2} (5-p)^{-{1\over 2}} \ell^{-{3\over 2}} 
r^{-{13-3p\over 4}},
\quad
f_-^x \sim 2 (5-p)^{-{1\over 2}} \ell^{1\over 2} 
r^{{(7-p)\over 4}},
\end{equation}
\begin{equation}
f_+^y \sim (2\ell)^{-{ 9-p\over 2(5-p)}}r^{-{7-p \over 4}},
\quad
f_-^y \sim (2\ell)^{ p-1\over 2(5-p)}r^{{p+1 \over 4}}.
\end{equation}
Thus, from (\ref{finalsope}), 
\begin{eqnarray}
S_x(T) &\sim& (|t_i-t_f|\Lambda)^{ -2\left[a^{\dagger}a
+ \frac{1}{2} \right]},
\\
S_y(T) &\sim& (|t_i-t_f|\Lambda)^{ -\frac{4}{5-p} \left[a^{\dagger}a+\frac{1}{2} \right]}.
\end{eqnarray}
The behavior of the $SO(7-p)$ directions is 
 consistent with the field-theory analysis for general 
$p$ in supergravity 
discussed in Appendix A. 2.   On the other hand, 
the behavior of the $SO(1+p)$ excitations 
conforms to the generalized BMN conjecture 
that they correspond to the derivatives $D_iZ$ of 
the field $Z$ 
along the base-space directions of D$p$-branes.

For $p\ne 3$ and $n\ne 0$, it is difficult to solve this equation exactly.   
By recalling  that the two-point S-matrix is governed by 
the asymptotic behaviors of the solutions $f_{\pm}$ for 
large $r$ or $T\rightarrow T_b$, 
we can try to extract 
some information for the solutions after making the large-$r$ approximation for the differential equations 
themselves as 
\EQ
\ell^2 r^{5-p} \frac{d^2 X}{dr^2} + \frac{5-p}{2} \ell^2 r^{4-p}   \frac{dX}{dr}  
-\left( m_{\rm app}^2 +{n^2 \over \alpha^2}\right)X=0
\label{approeq}, 
\EN
where
\EQ
m_{\rm app}^2 = {(7-p)(13-3p) \over 16} \ell^{2}r^{3-p}\; (=m_{{\rm app};p,x}^2),
\EN
or 
\EQ
m_{\rm app}^2 ={(7-p)(p+1) \over 16 } \ell^{2}r^{3-p}\; (=m_{{\rm app};p,y}^2),
\EN
for $SO(1+p)$ and $SO(7-p)$ directions, respectively. 
This is valid when 
\[
\ell^2r^{(5-p)} \gg 1, \quad {n^2 \over \alpha^2} \gg{1 \over r^2} .
\]

The equation (\ref{approeq}) for $p\ne 3$ can be converted 
into the modified Bessel equation
\[
w^2{d^2Y\over dw^2}+w{dY\over dw}-(\nu^2 
+ w^2)Y=0
,
\] with  
\EQ
\nu=\Big|{p-5\over p-3}\Big| \quad \mbox{or} \quad \nu =\Big|{2\over p-3}\Big| , 
\EN
for the $SO(1+p)$  and $SO(7-p)$ directions, respectively, 
by making the 
redefinition 
\EQ
X(\tau)=w^{1/2}Y(w), \quad w={2n\over |3-p|\ell \alpha}
r^{(p-3)/2} .
\EN 
Thus the asymptotic form of the solutions 
$f_{\pm}$ is described by the following 
general form
\EQ
X= C_1 r^{-{3-p \over 4}} I_{\nu}\left({2n\over |3-p|\ell \alpha}r^{-{3-p\over 2}}
\right)
+  C_2 r^{-{3-p\over 4}} K_{\nu}\left({2n\over |3-p|\ell \alpha}r^{-{3-p\over 2}}
\right).
\label{geneform}
\EN
The coefficients $C_1, C_2$ in general depend 
on $n/\alpha$ and $p$.  Since we have the exact solutions 
for $n=0$, these coefficients are determined 
by matching the above form in the limit 
$n\rightarrow0$. When $p=0$, for example, it gives \EQA
f^{{app};p=0,x}_+
&=& \sqrt{5}\, 3^{-{1\over 3}} \Gamma({2\over 3}) \left( {n\over \alpha}\right)^{-{5\over 3}}
\ell^{1/6}
r^{-{3\over 4}} I_{5\over 3}\!\left({2n\over 3\ell \alpha}r^{-{3\over 2}}
\right)
 \, 
\sim {\sqrt{5}\over 10}
\ell^{-{3\over 2}} r^{-{13\over 4}} ,
\label{eq:f+np0x}
\\
f^{{app};p=0,x}_-
&=& {2 \sqrt{5} \, 3^{1\over 3} 
\over 15 \Gamma({2\over 3})  }
 \left( {n\over \alpha}\right)^{5\over 3} \ell^{-{7\over 6}}
r^{-3/4} K_{5\over 3}\!\left({2n\over 3\ell \alpha}r^{-{3\over 2}}
\right)
\nonumber\\
&& \qquad\qquad\qquad\qquad \, \, \, 
+ C \,r^{-{3\over 4}} I_{5\over 3}\!\left({2n\over 3\ell \alpha}r^{-{3\over 2}}\right)
\, 
\sim {2\sqrt{5}\over 5}
\ell^{{1\over 2}} r^{7\over 4}
\label{eq:f-np0x}
,
\EQN
for the 
$SO(1)$ direction. 
In the last line, a constant $C$ remains unfixed.
This can be fixed by using the next-to-leading 
contribution of $f_-$ for $n=0$.
However, we can set $C=0$ since 
only the leading behavior is important for our purpose.

In the general case with $p<3$, 
we can set the asymptotic forms for $\tau 
\rightarrow T_b$ $(r\rightarrow \infty)$ as 
\EQ
f_+(\tau) = c^+_{\nu} w^{1/2} I_{\nu}(w),
\quad 
f_-(\tau)=c^-_{\nu}w^{1/2} K_{\nu}(w)
.
\label{geneasympform}
\EN
The Wronskian normalization condition 
requires 
\EQ
c^+_{\nu}c^-_{\nu}=\alpha/|n|. 
\EN
If we  interpolate these asymptotic solutions 
deep inside the bulk and directly 
impose the continuity condition at $\tau=0$, 
these undertermined coefficients $c_{\nu}^{\pm}$ 
are fixed by requiring the condition 
of time-reflection symmetry. In the limit $n\rightarrow 0$,  
this can indeed be done as exemplified 
above for $p=0$, and the resulting general 
asymptotic form of the 
coefficients is  
\EQ
\lim_{n\rightarrow 0}
c^+_{\nu} \rightarrow 
\sqrt{\alpha\nu\over 2|n|}2^{-(3-p)\nu/(5-p)}
\Gamma(\nu) (p-3)^{\nu}\Big|{n\over \alpha}\Big|^{-\nu}
\ell^{2\nu/(5-p)}
.
\EN
For general $n$, the symmetry under the 
scaling (\ref{scaling}) demands that  
the coefficients take the form 
\EQ
c^+_{\nu}=\sqrt{{\alpha\over |n|}}
c_{\nu}(s), 
\quad 
c^-_{\nu}=\sqrt{{\alpha\over |n|}}
\Big(c_{\nu}(s)\Big)^{-1}
.
\EN
Here the argument in the undertermined function 
$c_{\nu}(s)$ is given as 
\EQ
 s\equiv {|n|\over \alpha}\ell^{-2/(5-p)}=
|n|q_p^{1/2}\ell^{(3-p)/(5-p)}/J
\label{s}
\EN
in terms of the angular momentum $J$ and the 
distance $\ell =(5-p)|t_f-t_i|/4$.  
 Similarly, the argument of the modified Bessel 
function at the cutoff scale is given by
\EQ
w={2|n|\over (3-p)\ell\alpha}r(T)^{(p-3)/2}
={2\over 3-p}\left({2\over 5-p}\right)^{(p-3)/(5-p)}
{|n|q_{p}^{1/2}\over J}\Lambda^{(p-3)/(5-p)}=
{|n|\over \alpha}(T_{b}-T), 
\label{w}
\EN
which contains the cutoff parameter $\Lambda$,  but 
is actually {\it independent of} the target-space distance $\ell$. 
We note that the appearance of these new variables is 
basically due to the existence of new scale $q_p^{1/2}$, 
playing the role of effective string tension. 
Remember that $q_p^{1/2}$
appears as the over-all prefactor of the metric 
tensor.

We can understand why the cutoff-dependent variable  $w$ 
is associated with the 
stringy excitations as follows: 
First, recall that the effective scale
of the background metric
 is of order $R_c \equiv 
\sqrt{q_p^{1/2}r^{-(3-p)/2}}$. Near the boundary 
this is of order $
\sim q_p^{1/(7-p)}$, since $r \sim q_p^{1/(7-p)}$. The 
contribution of Kaluza-Klein modes to the squared 
mass $M^2$ is then 
of order $M^2_{{\rm KK}} \sim (J/R_c)^2$, while the contributions of the 
stringy excitations is of order  $M_{st}^2 
\sim n$.  In terms of $M^2$, the generalized conformal 
dimensions of order $O(J)$ take the form
\[
\Delta_I \sim {M_{{\rm KK}}^2 \over (J/R^2_c)} \sim 
J + O(1). 
\] 
The corrections by the stringy excitations are 
 therefore expected to enter through the form 
$
{M_{st}^2\over (J/R^2_c)}\sim  {n R_c^2\over J} \sim w 
$
at least in the  regime where the background curvature is regarded as small. 

From these facts, we can see that, firstly,  
the cutoff dependence always enters in the 
S-operator in a factorized form. Thus, the cutoff 
dependence 
can be removed (or renormalized) by assuming a 
suitable normalization condition 
for 2-pt correlation functions, just as we have 
already seen for supergravity modes ($n=0$) without 
stringy excitations. Secondly, 
the expression (\ref{s}) 
implies that the long-distance behavior of the 
asymptotic solutions for $p<3$ and hence of the 
S-operator with respect to the distance $\ell
\rightarrow \infty$ (with fixed $n$) at the 
boundary is governed by  the large $|n|q_p^{1/2}/J$ behavior 
of the solutions.  Namely, we can study the 
infra-red properties of the dual gauge theories 
of D$p$-brane ($p<3$) by examining the 
limit of sufficiently large $n$ for fixed $\ell$ such that 
\EQ
n^2q_p/J^2 \gg \ell^{-2(3-p)/(5-p)}
. 
\label{irlimit}
\EN

Therefore let us  consider the behavior of the 
solutions  in the large 
$n$ limit. Since the cutoff of $r$ is of order $q_p^{1/(7-p)}$, 
we can neglect the mass term arising from the 
background comparing with the mass corresponding to 
the stringy excitation, if we consider the limit of  extremely  
large $n^2q_p/J^2$  such that 
\[
{n^2\over \alpha^2} ={n^2\ell^2 q_p\over J^2} 
\gg \ell^2 q_p^{(3-p)/(7-p)}\rightarrow 
n^2q_p/J^2\gg q_p^{(3-p)/(7-p)}
.
\]
Though this (cutoff dependent) condition is different from the above 
condition (\ref{irlimit}) which is sufficient 
for the infra-red behavior, 
the scaling symmetry allows us to
 assume that no new scale enters in the 
computation of the coefficient function $c_{\nu}(s)$. 
  Then, the original differential equation 
is reduced to the trivial one $[{d^{2}\over d\tau^{2}}-({n\over \alpha})^{2}]X=0$, and thus the solutions 
are now approximated by 
\EQA
f_{+}&=&\sqrt{\alpha\over 2|n|}e^{-{|n|\over \alpha}\tau},\\
f_{-}&=&\sqrt{\alpha\over 2|n|}e^{{|n|\over \alpha}\tau}, 
\label{eq:flatsol}
\EQN
which should be matched with the large $n$ limit of 
(\ref{geneasympform}). Using the last of various equivalent 
expressions for $w$  in (\ref{w}) and the 
asymptotic form of the modified 
Bessel functions $(I_{\nu}(w), 
K_{\nu}(w)) 
\rightarrow (\e^w/\sqrt{2\pi w}, 
\e^{-w}/\sqrt{2w/\pi}) $, we obtain the 
asymptotic form of the function $c_{\nu}(s)$ 
as 
\EQ
c_{\nu}(s)\sim \sqrt{\pi}\e^{-|n|T_b/\alpha}
\EN
for large $n$.  Since $T_b \sim \ell^{-2/(5-p)}$ 
for $p<3$, the argument in the exponential 
is proportional to $s$ as it should be.   This shows that the 
contribution to the S-operator from the  
stringy excitations with nonzero $n$ in the large distance limit 
of the boundary theory is of the form
\EQ
S(T)_{p<3} \sim \exp\{-{\hat{c}|n|q_{p}^{1/2}\over J}|t_{f}-t_{i}|^{(3-p)/(5-p)}
(a^\dagger a+1/2)\}
\EN
where $\hat{c}$ is a numerical constant. 
This two-point function is exponentially damped compared with 
the non-stringy modes.
In other words, the 
infra-red limits of the dual gauge theories 
for $p<3$ are described  by non-stringy 
supergravity modes, and the stringy modes are 
decoupled. 

If we repeat the above 
computation in the case $p=3$, we get the 
correct 2-pt function for the large $n$ limit, 
$S \sim \left( {1\over |t_{f}-t_{i}| \Lambda}
\right)^{{2q_{p}^{1/2}|n|\over J}(a^{\dagger}a+1/2)}$, 
which is of course power-behaved with anomalous 
conformal dimensions 
($\sqrt{1+(n^2R^4/J^2)} 
\sim q_{p}^{1/2}|n|/ J
$) and hence the stringy 
modes cannot be neglected in the 
infra-red limit for $p=3$. In contrast to this,  the stringy excitations for $p<3$ cannot be described simply by the 
shift of generalized conformal dimension to 
anomalous generalized conformal dimensions. 

Let us briefly touch on 
the opposite short-distance behavior of 2-pt 
functions for $p<3$. Clearly, we expect that 
the short-distance structure 
is encoded in the small $s$ behavior of the 
function $c_{\nu}(s)$. Because 
the cutoff $\Lambda$ associated with $r$  
is expected to play the role for the short distance 
cutoff also for the boundary theory 
as $\ell > 1/\Lambda\sim q_p^{-(5-p)/2(7-p)}$, 
$s$ near the UV cutoff behaves as $s\sim |n|q_p^{1/2}q_p^{-(3-p)/2(7-p)}/J
\sim  |n| q_p^{2/(7-p)}/J$. Therefore, for nonzero 
$n$, the asymptotic form for small $n$ can be 
used for the  short distance behavior 
when 
\EQ
|n| q_p^{2/(7-p)}\ll J
. 
\EN
It would be an interesting problem to study the 
function $c_{\nu}(s)$ for small $s$, beyond 
its asymptotic form 
given above,  by solving the 
differential equation perturbatively in $n^2/\alpha^2$. 
In the intermediate regime, the behavior of two-point 
functions are thus very nontrivial. 
 
Finally, let us consider the case $p=4$. 
One of the marked differences of this case from the cases 
$p<3$ is that the affine time interval 
from the boundary to boundary is infinite, 
as in the conformal case $p=3$. 
So near the boundary, we have 
\EQ
w={4|n|q_p^{1/2}\over J}\Lambda={|n|\over \alpha}\tau \rightarrow \infty, 
\EN
in contrast to $w\rightarrow 0$ for $p<3$. Also, the mass 
functions arising from the background vanishes asymptotically 
near the boundary.  Using these properties, we derive the asymptotic solutions for the stringy 
modes $n\ne 0$ as 
\EQ
f_+ \rightarrow \sqrt{{\alpha\over \pi |n|}}e^{-w}, \quad 
f_-\rightarrow \sqrt{{\pi \alpha\over |n|}}e^{w}. 
\EN
It is easy to check that these asymptotic behaviors  can be 
matched with the asymptotic forms obtained from the 
modified Bessel equation. The difference from the cases $p<3$ 
is that we have to replace the roles of 
$I_{\nu}(w)$ and $K_{\nu}(w)$.  

The above form of the asymptotic solutions implies that the contributions 
$(2f_-\dot{f}_-)^{-(a^{\dagger}a+{1\over 2})}$
of the stringy modes to the S-operator does not contain 
any $\ell$ dependence, and hence does not modify 
the behaviors of 2-pt correlation functions 
determined by the lowest supergravity modes with $n=0$. 
Thus, the 2-point functions of the stringy BMN operators 
of the boundary theory, (4+1)-dimensional Yang-Mills 
theory with maximal supersymmetry, 
are completely 
degenerate with those of supergravity operators without 
stringy excitations. This is almost the free-field behavior, 
but is not quite so,  since the contributions of supergravity 
modes are 
\EQ
S_x(T) \sim \ell^{-(2a^{\dagger}a +1)} , \quad 
S_y(T) \sim  \ell^{-4(a^{\dagger}a + {1\over 2})}
. 
\EN
The free-field behavior would correspond to 
$S_y(T) \sim \ell^{-3(a^{\dagger}a + {1\over 2})}$.
This strongly suggests that the infra-red behavior 
of this system is governed by a nontrivial fixed point. 
The degeneracy will be lifted when we take into account 
the interactions of various stringy modes including 
higher-loop effects, since 
the string vertices have complicated dependence on the 
external string states.

\section{Conclusion}
To summarize, we have discussed how the 
tunneling picture proposed in ref. \cite{dobashimayone} 
is utilized to extend the holographic 
interpretation of the PP-wave limit to 
general nonconformal backgrounds of D$p$-branes. 
We have then developed a general quantum theory with 
time dependent masses, in order to extract 
predictions for two-point correlation functions 
of the holographically dual gauge theories. 
The behaviors of the resulting two-point 
functions for non-stringy supergravity modes 
are consistent with available 
field-theoretical analyses using supergravity. 
The behavior of the stringy BMN states turned out to be very 
different from that of D3-branes, 
though roughly the behaviors are not 
in contradiction with the conventional expectations 
for superrenormalizable ($p<3$) and 
nonrenormalizable ($p=4$) gauge theories.  
The implications we found for the dual gauge theories 
seem to be important, since for all $p<5 \, \, (p\ne 3)$ the 
infra-red structure is very nontrivial. In particular, 
we pointed out that our results may be 
interpreted as a strong indication for the 
existence of nontrivial fixed points 
in non-conformal super Yang-Mills theories 
of D$p$-branes.  

We hope that our work 
laid a foundation for further investigations 
of holography for general D$p$ branes 
using the PP-wave limit.  
Apart from an immediate extension of the 
present formalism to include fermionic excitations,  
there are many directions of further researches 
along the line of our work, such as 
treatments  of higher-point correlation functions 
and string-loop effects, considerations of 
other possible classical backgrounds 
including spinning strings \cite{gkp2}\cite{tseytlin} and 
extension to 11-dimensional theories in connection with 
supermembranes.  

Another important task 
is of course to develop nonperturbative methods 
for studying correlation functions directly 
within the framework of  holographically dual 
gauge theories themselves. 
Some sort of generalized mean-field approaches, extending 
those pursued in \cite{kabatetal}\cite{suginokawai}, 
seems to be 
a promising direction, in view of the fact 
that the infra-red behaviors of the systems  
are predicted to be almost free-field like 
for $p=4$, while for $p<3$ the corrections 
to free-field behaviors with respect to 
stringy modes are sufficiently dramatic  to decouple 
the stringy excitations in the infra-red
 from the supergravity modes. 
Note also that the powers exhibited 
in 2-pt S-operators with respect to the 
distance $\ell$ are different from the 
free field behavior in all the cases except for 
the conformal case $p=3$. 

The shifts  
of exponents from those expected from 
free-field theories should correspond, at least in 
some qualitative sense,   to the 
presence of nontrivial mean fields for 
$p\ne 3$. It is of some interest 
 to define effective 
dimensionality from the shifted exponents. 
Since free massless theories with $d_{{\rm eff}}$ base-space 
dimensions would  behave as $\ell^{-(d_{{\rm eff}}-2)}$, the $SO(7-p)$ 
S-operator implies 
\[
d_{{\rm eff}}=2+ {4\over 5-p}. 
\]
For $p=1$ and $p=4$, remarkably, we have integer effective 
dimensions $d_{{\rm eff}}=3$ and $d_{{\rm eff}}=6$, 
respectively, which 
are related to the true base-space dimensions $d$ 
by $d_{{\rm eff}}=d+1$. It would be an interesting dynamical 
question whether this phenomenon can be 
interpreted from the viewpoint of M-theory 
as being related to M2 and M5 branes, respectively. 
For instance,  
it has been shown in a previous work by two of the present authors \cite{sekiyone2} 
that the supermembrane 
and IIA matrix-string theory \cite{matstring} 
in the large $N$ limit 
can be directly related by a new 
particular matrix regularization. 
From the viewpoint of the matrix-string, the 
Yang-Mills coupling constant is reversed. Therefore, 
the weak coupling limit 
$g_s\sim g_{{\rm YM}}^2 \rightarrow 0$  in the context of 
the present work 
must correspond to the M-theory limit $R_{11}
(\sim 1/\sqrt{g_{{\rm YM}}}) \rightarrow 
\infty$ by which the 11th dimension is de-compactified.  In this sense, the above result 
$d_{{\rm eff}}=3$ for $p=1$  may have a natural 
interpretation.  Also, in the case $p=4$, it would be very interesting if $d_{{\rm eff}}=6$ is related to the argument in \cite{seiberg} 
on  the relevance of the 6-dimensional  fixed-point 
theory with (2,0) supersymmtry for the infra-red limit of 5-dimensional  
maximally supersymmetric Yang-Mills theory.

%\newpage
\vspace{0.8cm}
\noindent
Acknowledgements

The work of T. Y. is supported in part by Grant-in-Aid for Scientific Research (No. 12440060 and No. 13135205)  from the Ministry of  Education, Science and Culture. 
The work of M.~A. is supported in part by the fellowship from Soryushi Shogakukai. The work of Y.~S. is supported in part by the Research Fellowships of the Japan Society for the Promotion of Science
for Young Scientists. M.~A. and Y.~S. would like to thank Edward Witten 
for a valuable discussion at KEK. 

%\newpage

\vspace{0.8cm}
\noindent
\appendix

\section{Supergravity analysis of two-point functions}
\setcounter{equation}{0}
\renewcommand{\theequation}{\Alph{section}.\arabic{equation}}
\renewcommand{\thesubsection}{\Alph{section}.\arabic{subsection}}

In this Appendix, we summarize briefly 
the calculation of the
two-point functions  based on
the more standard supergravity analysis. 
We apply the general prescription proposed 
by Gubser, Klebanov and Polyakov and by Witten~\cite{holography}
to the non-conformal theories.
In {\it A.1}, we review the calculation 
for the D0-brane background performed in the previous works \cite{sekiyone}.
In {\it A.2}, we discuss the form of the two-point function 
of the operators corresponding to supergravity modes
with large angular momentum for the case of general D$p$-branes.

\subsection{Two-point functions for the D0-branes}

In \cite{sekiyone}, complete spectrum of the supergravity
fluctuations around the near-horizon 
D0-brane background has been worked out. By expanding 
the fields into the spherical harmonics on ${\rm S}^{8}$,
and diagonalizing the linearized equations of motion,
it was found that each physical mode 
of the bosonic fluctuations is described by
the following scalar action in the $(t,z)$ space\footnote{Note 
that the normalization of 
the coordinates $t$ and $z$ used in this paper differs 
from the conventional one used in \cite{sekiyone} by 
dimensionful factor. The relation
is $(t,z)|_{here}=q^{-1/2}(t,z)|_{conventional}$.}:
\begin{equation}
S={1\over 8\kappa^{2}}Cq^{3/2} \int dt dz z
\left\{ (\partial_{0}\psi_{I,\ell_{I}})^{2}
%\partial_{0}\psi_{I,\ell_{I}}
+(\partial_{z}\psi_{I,\ell_{I}})^{2}
%\partial_{z}\psi_{I,\ell_{I}}
+{\nu_{I,\ell_{I}}^{2}\over z^{2}}\psi_{I,\ell_{I}}^{2}\right\}.
\label{eq:sgaction}
\end{equation}
where we are considering the Euclidean space-time 
and $q_0$ is replaced by $q$ for notational brevity. 
$C$ is a numerical constant and $\kappa^{2}=g_{s}\ell_{s}^{8}$
is the Newton constant in 10 dimensions. We have denoted
the field collectively as $\psi_{I,\ell_{I}}$,
where $I$ labels the modes and, $\ell_{I}$ is the total 
angular momentum on ${\rm S}^{8}$. The constant $\nu_{I,\ell_{I}}$
is given by $\nu_{I,\ell_{I}}=2\ell_{I}/5+c_{I}$, where $c_{I}$
is determined by the explicit diagonalization. 
For example, for the traceless symmetric 
tensor modes on ${\rm S}^{8}$, we have $c_{I}=7/5$. 
See \cite{sekiyone} for the spectrum of $\nu_{I,\ell_{I}}$.
(In the following, we will suppress the subscript $(I,\ell_{I})$ 
for brevity.)

The equation of motion for the Fourier mode
$\psi_{\omega}$ ($\psi(t,z)=\int d\omega e^{i\omega t}
\psi_{\omega}(z)/2\pi$)
is 
\[
\left[
\partial_{z}^{2}+{1\over z}\partial_{z}
-(\omega^{2}+{\nu^{2}\over z^{2}})\right] \psi_{\omega}(z)=0.
%\label{eq:sgeom}
\]
This  is solved by the modified Bessel function.
We assume that the solution is regular at the origin 
($z\to \infty$: $r\to 0$), and we impose
a boundary condition at the end of the near-horizon region
($z=q^{-5/14}\to 0$: $r\sim q^{1/7}\to \infty$):
\EQ
\psi_{\omega}(q^{-5/14})=\lambda_{\omega}.
\label{eq:sugrabc}
\EN
The solution which satisfy these conditions is written as
\EQ
\psi_{\omega}(z)=\lambda_{\omega}{K_{\nu}(z|\omega|)\over 
K_{\nu}(q^{-5/14}|\omega|)}.
\label{eq:sgsol}
\EN

We assume the relation between the classical supergravity action
and the generating functional of the gauge-theory correlators,
in the form with cutoff as proposed by Gubser, Klebanov and Polyakov:
\EQ
e^{-S_{cl}[g]}|_{\psi(q^{-5/14},t)=\lambda(t)}
=\langle e^{\int dt \lambda(t) {\cal O}(t)}\rangle.
\EN
Here, $S_{cl}$ is the classical value of the
supergravity action which is a functional of the
boundary value of the field.
By evaluating the action 
(\ref{eq:sgaction}) with the classical solution
(\ref{eq:sgsol}), we obtain the two-point function
\EQA
\langle {\cal O}(t_{1}) {\cal O}(t_{2})\rangle
&=& \int {d\omega_{1}\over 2\pi}\int {d\omega_{2}\over 2\pi}
e^{i\omega_{1}t_{1}}e^{i\omega_{2}t_{2}}
\langle {\cal O}(\omega_{1}) {\cal O}(\omega_{2})\rangle
\nonumber\\
&=&-\int {d\omega_{1}\over 2\pi}\int {d\omega_{2}\over 2\pi}
e^{i\omega_{1}t_{1}}e^{i\omega_{2}t_{2}}
(2\pi){\delta\over \delta \lambda_{\omega_{1}}}
(2\pi){\delta\over \delta \lambda_{\omega_{2}}}
S_{cl}[\lambda]|_{\lambda=0}\nonumber\\
&=&{C\over 8\kappa^{2}}q^{3/2}\int d\omega e^{i\omega(t_{1}-t_{2})}
q^{-5/14}{\partial_{z}K_{\nu}(z|\omega|)|_{z=q^{-5/14}}
\over K_{\nu}(q^{-5/14}|\omega|)}.\nonumber
%\label{eq:tp1}
\EQN

%Let us evaluate the supergravity action (\ref{eq:sgaction})
%with this solution. The bulk contribution vanish due to the
%equation of motion, and we have only the boundary term
%\EQA
%S_{cl}&=&{C\over 8\kappa^{2}}q\int {d\omega\over 2\pi}\left[
%zg_{\omega}\partial_{z}g_{-\omega}\right]_{z=q^1/7}^{z=\infty}
%\nonumber\\
%&=&-{C\over 8\kappa^{2}}q\int {d\omega\over 2\pi}
%\lambda_{\omega}\lambda_{-\omega}q^{1/7}
%{\partial_{z}K_{\nu}(z|\omega|)|_{z=q^{1/7}}
%\over K_{\nu}(q^{1/7}|\omega|)}
%\EQN

In order to study the behavior in the region of large
$|t_{1}-t_{2}|$, we expand the integrand in powers of
$|\omega|$. In the case of $\nu\neq$ integer, we have
%\EQA
%K_{\nu}(|\omega|z)&=&
%{\pi\over 2}{1\over \sin \nu\pi }\left( {|\omega|z\over 2}
%\right)^{-\nu}{1\over \Gamma (-\nu +1)}
%\Big[ 1+ \mbox{(positive integer power of $(|\omega|z)^{2}$)}
%\nonumber\\
%&&\qquad - {\Gamma(-\nu+1)\over \Gamma(\nu+1)}
%\left({|\omega|z\over 2} \right)^{2\nu} \left\{
%1+ \mbox{(p.i.p. of $(|\omega|z)^{2}$)}\right\}\Big].
%\label{eq:besselex1}
%\EQN
%\EQAN
%\partial_{z}K_{\nu}(|\omega|z)&=&
%{\pi\over 2}{1\over \sin \nu\pi }\left( {|\omega|z\over 2}
%\right)^{-\nu}{1\over \Gamma (-\nu +1)}{1\over z}
%\Big[ -\nu + \mbox{(p.i.p. of $(|\omega|z)^{2}$)}
%\nonumber\\
%&&\qquad - \nu {\Gamma(-\nu+1)\over \Gamma(\nu+1)}
%\left({|\omega|z\over 2} \right)^{2\nu} \left\{
%1+ \mbox{(p.i.p. of $(|\omega|z)^{2}$)}\right\}\Big].
%\EQNN
\EQA
\langle {\cal O}(t_{1}) {\cal O}(t_{2})\rangle
&=&{C\over 8\kappa^{2}}q^{8/7}
\int d\omega e^{i\omega (t_{1}-t_{2})}
\Big[ -\nu + 
%\mbox{(p.i.p. of $(|\omega|q^{-5/14})^{2}$)}
\cdots
\nonumber\\
&&\qquad 
-{2\nu\over q^{-5/14}}{\Gamma(-\nu+1)\over \Gamma(\nu+1)}
\left( {|\omega|q^{-5/14}\over 2}\right)^{2\nu} 
(1+
%\mbox{(p.i.p. of $(|\omega|q^{-5/14})^{2}$)})\Big], \nonumber
\cdots)\Big]
%\label{eq:intgtwopoint}
\EQN
where the dots represents the terms which are of
positive integer powers in $(|\omega|q^{-5/14})^{2}$.

The Fourier integral in the above expression 
is performed using the formula
\[
\int_{-\infty}^{\infty} 
d\omega e^{-i\omega(t_{1}-t_{2})}|\omega|^{2\nu}
={\Gamma(2\nu+{1\over 2}) \over \Gamma(-\nu)}
{2^{\nu+1}\sqrt\pi \over |t_{1}-t_{2}|^{2\nu+1}},
\]
which is valid for $\nu\neq 0,1,2,\ldots, -{1\over 2},-{3\over 2},
\ldots$. Integral of the terms analytic in $\omega$ are divergent,
and correspond to the derivatives of the delta function 
$\delta(t_{1}-t_{2})$. Ignoring these divergences,
the leading term of the correlator 
when $|t_{1}-t_{2}|\gg q^{-5/14}$ is\footnote{
Two-point function in the conventional normalization of 
$t$
%, in which the world-volume coordinates on 
%the D-branes have length dimension 1, 
is obtained by the rescaling $t\to q^{-1/2}t$
and ${\cal O}\to q^{1/2}{\cal O}$ (since we are
considering the operators with length dimension $-1$):
\[
\langle {\cal O}(t_{1}) {\cal O}(t_{2})\rangle|_{conventional}
={C\over 8\kappa^{2}}2^{-\nu+2}\sqrt{\pi}
{\Gamma(2\nu+{1\over 2})\over \Gamma(\nu -1)}
{q^{1+2\nu/7}\over |t_{1}-t_{2}|^{2\nu+1}}.
%\label{eq:gtwopoint}
\]}
\EQ
\langle {\cal O}(t_{1}) {\cal O}(t_{2})\rangle
={C\over 8\kappa^{2}}2^{-\nu+2}\sqrt{\pi}
{\Gamma(2\nu+{1\over 2})\over \Gamma(\nu -1)}
{q^{3/2-5\nu/7}\over |t_{1}-t_{2}|^{2\nu+1}}.
\label{eq:gtwopoint}
\EN
The right hand side has scaling dimension $\Delta=-1+10\nu/7$ 
with respect to the generalized conformal symmetry.
For each supergravity modes, we can consistently 
determine the corresponding operators ${\cal O}$ which have
the correct $\Delta$, 
as shown in the tables in \cite{sekiyone}. 
The scaling dimension 
of an operator is 
determined solely from 
its 11-dimensional tensor structure, and is given by (4.23) in the text.

\subsection{Large $J$ behavior of the two-point functions 
for the D$p$-branes}

To our knowledge, supergravity spectrum 
on D$p$-brane backgrounds for $p\neq 0,3$ has not been 
analyzed previously, except for the partial results for $p=1$ 
reported in \cite{D1spec}.
Instead of deriving the precise correspondence 
between gauge theory operators
and supergravity modes, which of course 
deserves for a separate work, 
 we discuss here the form of the correlators
in the large angular momentum limit 
by an illustrative calculation.

%We present a simplified version of 
%the calculation of two-point function from supergravity,
%in the cases of the general D$p$-branes.
%The purpose is to
%discuss the $J$ (angular momentum) dependence of the correlators
%in the large $J$ limit.

Let us consider a massless scalar field $\varphi$ 
which couple to the metric and the dilaton
of D$p$-brane background
\EQ
S={1\over 8\kappa^{2}}\int d^{10}x \sqrt{-g}e^{-2\phi}
g^{\mu\nu}\partial_{\mu}\varphi\partial_{\nu}\varphi,
\label{eq:Dpscalar}
\EN
and calculate the gauge theory two-point function,
by applying the GKP-Witten prescription.
Strictly speaking, this action simply for a usual 
scalar field $\phi$ 
may not really describe the fluctuations
around the D$p$-brane (except for $p=3$, where the
dilaton background is constant), but the dependence on
$J$ will be inferred from this analysis.

Substituting the D$p$-brane metric (\ref{eq:Dpmetric2}) and the dilaton
$e^{\phi}=q_{p}^{(3-p)/4}e^{\tilde{\phi}}$ with $e^{\tilde{\phi}}$
given in (\ref{etildephi}), 
the action becomes 
\[
S={\tilde{C}_{1}q_{p}^{(p+1)/2}\over 8\kappa^{2}}
\int d^{10}x  z^{9-p\over p-5}
\sqrt{\tilde{g}}\left\{
\delta_{mn}\partial_{m}\varphi\partial_{n}\varphi+
\partial_{z}\varphi\partial_{z}\varphi
+\left({2\over 5-p}\right)^{2}{1\over z^{2}}\tilde{g}^{ij}
\partial_{i}\varphi\partial_{j}\varphi\right\},
\]
where the base space of the gauge theory $x^{m}$ $(m,n=0,1,\ldots,p)$
is Euclidean, and $\tilde{g}_{ij}$ 
is the metric of unit ${\rm S}^{8-p}$.
$\tilde{C}_{1}$ and $\tilde{C}_{2}$, $\tilde{C}_{3}$ below
are numerical constants.
Note that when $\varphi$ is a spherical harmonics on 
${\rm S}^{8}$ with 
angular momentum $J$, the last term becomes
\[
\int d^{8-p}\tilde{x}\sqrt{\tilde{g}}\tilde{g}^{ij}
\partial_{i}\varphi\partial_{j}\varphi=
\int d^{8-p}\tilde{x}\sqrt{\tilde{g}}
J(J+7-p)\varphi^{2}
\]
where $\tilde{x}_{i}$ are the coordinate on ${\rm S}^{8-p}$.
By performing a field redefinition 
$\varphi=q_{p}^{1/2}z^{(7-p)/(5-p)}\hat{\varphi}$,
and ignoring total derivative, the action becomes
\EQ
S={\tilde{C}_{2}\over 8\kappa^{2}}q_{p}^{(p+3)/2}\int d^{p+1}x dz z
\left\{
\delta_{mn}\partial_{m}\hat{\varphi}\partial_{n}\hat{\varphi}+
\partial_{z}\hat{\varphi}\partial_{z}\hat{\varphi}
+{\nu^{2}\over z^{2}}\hat{\varphi}^{2}\right\},
\label{eq:Dpaction}
\EN
where  
\[
\nu^{2}=\left({2\over 5-p}\right)^{2}J(J+7-p)+\left({7-p\over 5-p}\right).
\]
In the large $J$ limit, we have
\EQ
\nu\sim {2\over 5-p}J.
\label{eq:nulargej}
\EN
It seems reasonable to assume that this leading behavior 
aside perhaps from the constant part is
true for all the physical modes on the D$p$-background,
as we have confirmed in the case of the D0-brane background~\cite{sekiyone}.

We calculate the two-point function by evaluating the
supergravity action classically, in the same way
as for the D0-branes.
We expand the fields into Fourier mode $\hat{\varphi}_{k}$
\[
\hat{\varphi}(z,x^{m})=\int {d^{p+1}k\over (2\pi)^{p+1}}e^{ik_{m} x_{m}}
\hat{\varphi}_{k}(z),
\]
and impose the boundary condition at the end of the near-horizon region
$(z=q_{p}^{-{(5-p)/(14-2p)}})$.
%\EQ
%\left[\partial_{z}^{2}+{1\over z}\partial_{z}-(k^{2}+{\nu^{2}\over z^{2}})
%\right] \hat{\varphi}_{k}(z)=0.
%\EN
We take the solution of the equation of motion
\EQ
\hat{\varphi}_{k}(z)=\lambda_{k}{K_{\nu}(z|k|)\over 
K_{\nu}(q_{p}^{-(5-p)/(14-2p)}|k|)},
\EN
and evaluate the action (\ref{eq:Dpaction}) classically.
%\EQA
%S&=&{C\over 8\kappa^{2}}Q_{p}\int {d^{p+1}k\over (2\pi)^{p+1}}
%\Big[z\hat{\varphi}_{k}\partial_{z}\hat{\varphi}_{-k}
%\Big]_{z=Q^{1\over 7-p}}^{z=\infty}\nonumber\\
%&=&-{C\over 8\kappa^{2}}Q_{p}\int {d^{p+1}k\over (2\pi)^{p+1}}
%\lambda_{k}\lambda_{-k}Q_{p}^{1\over 7-p}
%{\partial_{z}K_{\nu}(z|k|)|_{z=Q^{1\over 7-p}}
%\over 
%K_{\nu}(Q^{1\over 7-p}|k|)}
%\EQN
Following the GKP-Witten prescription
\EQ
e^{-S_{cl}[g]}|_{\psi(q_{p}^{-{(5-p)/ (14-2p)}},x)=\lambda(x)}
=\langle e^{\int d^{p+1}x \lambda(x) {\cal O}(x)}\rangle,
\EN
we obtain the two-point function 
\EQA
\langle {\cal O}(x){\cal O}(x')\rangle
%&=&-\int {d^{p+1}k\over (2\pi)^{p+1}}\int {d^{p+1}k'\over (2\pi)^{p+1}}
%e^{ik_{m}x_{m}}e^{ik'_{m}x'_{m}}
%(2\pi)^{p+1}{\delta\over \delta \lambda_{k}}
%(2\pi)^{p+1}{\delta\over \delta \lambda_{k'}}
%S[\lambda]|_{\lambda=0}\nonumber\\
&\!=\!&{\tilde{C}_{2}
\over 8\kappa^{2}}q_{p}^{(p+3)/2}\int {d^{p+1}k\over (2\pi)^{p+1}}
e^{ik_{m} (x_{m}-x'_{m})}
q_{p}^{-(5-p)/(14-2p)}
{\partial_{z}K_{\nu}(z|k|)|_{z=q_{p}^{-(5-p)/(14-2p)}}
\over K_{\nu}(q_{p}^{-(5-p)/(14-2p)}|k|)}\nonumber\\
&& ={\tilde{C}_{2}\over 8\kappa^{2}}q_{p}^{(p+3)/2}
\int d^{p+1}k e^{ik_{m} (x_{m}-x'_{m})}q_{p}^{-(5-p)/(14-2p)}
\Big[ -\nu + 
\cdots
%\mbox{(p.i.p. of $(k^{2}q_{p}^{-(5-p)/(14-2p)})$)}
\nonumber\\
&& \quad
-{2\nu\over q_{p}^{-(5-p)/(14-2p)}}{\Gamma(-\nu+1)\over \Gamma(\nu+1)}
\left( {|k|q_{p}^{-(5-p)/(14-2p)}\over 2}\right)^{2\nu} (1+
%\mbox{(p.i.p. of $(k^{2}q_{p}^{-(5-p)/(14-2p)})$)}
\cdots )\Big]. \nonumber
%\label{eq:Dptpexp}
\EQN
where the dots represents terms with positive integer
powers in $(k^{2}q_{p}^{-(5-p)/(14-2p)})$. 
(We have assumed $\nu\neq$ integer when expanding
the Bessel functions.)
The first line of the right hand side gives the delta function 
divergences,
which we ignore. Performing the $k$ integral,
the leading part of the two-point function becomes
%\EQA
%\langle {\cal O}(x){\cal O}(x')\rangle
%&=&-{\tilde{C}\over 8\kappa^{2}}Q_{p}^{1+{2\nu\over 7-p}}
%\int_{0}^{\infty} d|k| \int_{-1}^{1} d(\cos \theta)
%\int d\Omega_{p-1}e^{i|k||x-x'|\cos\theta}|k|^{p+2\nu}\nonumber\\
%&=&-{\tilde{C}\over 8\kappa^{2}}Q_{p}^{1+{2\nu\over 7-p}}
%\int_{0}^{\infty}
%d|k|d\Omega_{p-1}{|k|^{p+2\nu-1}\over i|x-x'|}
%\left( e^{i|k||x-x'|}-e^{-i|k||x-x'|}\right) 
%\EQN
%where $\tilde{C}$ is a constant which can be
%read off from the above equations.
%After the momentum integral, (I have not calculated $\hat{C}$
%which appears below, but,) this should become
\[
\langle {\cal O}(x){\cal O}(x')\rangle
={\tilde{C}_{3}\over 8\kappa^{2}}{q_{p}^{1+{2\nu\over 7-p}}
\over |x-x'|^{p+1+2\nu}}.
%\label{eq:sugratpDp}
\]
From (\ref{eq:nulargej}), we see that the 
correlator of the operator 
${\cal O}={\rm Tr} (X^{J})$ (with large $J$),
which would correspond to the ground state of the
string in the PP-wave limit take the form 
\EQ
\langle {\cal O}(x){\cal O}(x')\rangle
\sim q_{p}^{4J\over (7-p)(5-p)}
{1\over |x-x'|^{4J\over 5-p}}.
\EN
Also, when we excite one
'y-oscillator' in the particle
picture, the exponent
will increase by $4/(5-p)$. 
These results are consistent with those we find 
using the PP-wave analyis in the text.

\section{Particle amplitude on the `effective metric' of the D0-branes}

In this Appendix, we argue that 
the diagonalized supergravity 
fluctuations around the D0-branes can be
regarded alternatively as the fields propagating on an 
effective metric, which is the direct product of
AdS and sphere. We study the particle amplitude on 
that effective background.
Unfortunately, the description cannot 
be applied for stringy modes. 

The effective action of the supergravity fluctuations 
(\ref{eq:sgaction}) can be regarded as the s-wave
part of a massive scalar action in 
an effective metric $\hat{g}_{\mu\nu}$,
(without a dilaton coupling):
\begin{equation}
S={1\over 8\kappa^{2}}\int d^{10}x \sqrt{-\hat{g}}\left\{ 
\hat{g}^{\mu\nu}\partial_{\mu}\hat{\psi}_{I,\ell_{I}}
\partial_{\nu}\hat{\psi}_{I,\ell_{I}}
+\hat{m}_{I,\ell_{I}}^{2}\hat{\psi}_{I,\ell_{I}}^{2}\right\}.
\label{eq:actioneff}
\end{equation}
Here, $\hat{g}_{\mu\nu}$ is related to the
original D0-brane metric $g_{\mu\nu}$ by a Weyl transformation
$\hat{g}{}_{\mu\nu}=(q^{3/4}e^{\tilde{\phi}})^{-2/7}g_{\mu\nu}$
and is just the (Euclidean) ${\rm AdS_{2}}\times {\rm S}^{8}$ metric:
\begin{equation}
\hat{ds}{}^{2}=\hat{g}_{\mu\nu}dx^{\mu}dx^{\nu}
=q^{2\over 7}\left\{
\left({2\over 5}\right)^{2}{dt^{2}+dz^{2}\over z^{2}}
+d\psi^{2}+\sin^{2}\psi d\Omega_{7}^{2}\right\}.
\label{eq:effmetric}
\end{equation}
The scalar field $\hat{\psi}_{I,\ell_{I}}$ and mass 
$\hat{m}_{I,\ell_{I}}$ are defined by
\begin{equation}
\hat{\psi}_{I,\ell_{I}}=q^{5/28}z^{1/2}\psi_{I,\ell_{I}},
%\propto e^{{5/21}\tilde{\phi}}\psi_{A,J}
\quad
\hat{m}{}_{I,\ell_{I}}^{2}=\left({25\over 4}\nu_{I,\ell_{I}}^{2}
-{25\over 16}\right)q^{-{2\over 7}}.
\label{eq:hatpsi}
\end{equation}
where $e^{\tilde{\phi}}$ is given in (\ref{etildephi}).
This effective action is equivalent with the one 
given in the first reference in \cite{sekiyone}. 

We have ignored the total derivative terms
when rewriting (\ref{eq:sgaction}) in the form
(\ref{eq:actioneff}). When calculating the gauge theory
correlators, such terms contribute only to the `delta function' 
divergences, mentioned in {\it A.1}. Also, the field redefinition
(\ref{eq:hatpsi}) does not affect the two-point function.
Namely, if we impose the boundary condition to $\hat{\psi}$, 
we obtain
the same result for the leading part of the two-point function
(\ref{eq:gtwopoint}). 

Note that when the angular momentum $\ell_{I}=J$ is large,
the mass is given by 
$\hat{m}^{2}_{I,J}\sim J^{2}q^{-2/7}$. 
This allows us to regard $\hat{\psi}_{I,J}$ modes 
(with large $J$) alternatively as the higher partial waves 
of a scalar field $\hat{\psi}_{I,\ell_{I}=0}$ on 
the space-time (\ref{eq:effmetric}). 
Indeed, if we assume $\hat{\psi}_{I,J}$ is a spherical harmonics
with angular momentum $J$, the derivative along ${\rm S}^{8}$
gives 
\[
\int d^{10}x
\sqrt{-\hat{g}}\hat{g}^{ij}\partial_{i}\hat{\psi}_{I,J}
\partial_{j}\hat{\psi}_{I,J}
=\int d^{10}x \sqrt{-\hat{g}}q^{-2/7}J(J+7)\hat{\psi}_{I,J}^{2},
\]
which agrees with the mass term in the $J\to\infty$ limit.
This fact suggests that the effective metric 
(\ref{eq:effmetric})
has a meaning in the 10-dimensional sense, at least in this limit. We expect that there is also a  semi-classical  picture 
based on the particle on this effective metric.

 Let us therefore 
study the particle amplitude on our 
 effective D0-brane metric.
Since the space-time is of the AdS$\times$S form, the amplitude is
obtained in the similar way as for the D3-brane case discussed
in the text. To obtain the classical contribution, 
consider the $(t,z,\psi)$ part of the massless particle action 
on (\ref{eq:effmetric}):
\begin{equation}
S={1\over 2}\int d\tau {1\over \eta}\left\{ {R^{2}\over z^{2}}
(\dot{t}^{2}+\dot{z}^{2})
-\left({5\over 2}\right)^{2}R^{2}\dot{\psi}^{2}\right\}
,
\end{equation}
where $R\equiv 2q^{1/7}/5$, and the double Wick-rotation
has been performed. 
Conjugate momentum for $\psi$ is 
%$J=-\left({5\over 2}\right)^{2}R^{2}{\dot{\psi}\over N}$,
$J=-(5R/2)^{2}\dot{\psi}/\eta$, and the Routh function becomes
\[
\bar{S}
={1\over 2}\int d\tau \{{1\over \eta}{R^{2}\over z^{2}}
(\dot{t}^{2}+\dot{z}^{2})
+\left({2\over 5}\right)^{2}{\eta\over R^{2}}J^{2}\},
\]
which is identical to the one for the D3-brane case if we
replace $J\to 2J/5$ in the latter. 
Proceeding similarly to that case,
we then obtain  
\begin{equation}
\langle t_f,1/\Lambda,J;T| t_i,1/\Lambda,J;-T\rangle_{class}=e^{-\bar{S}}
=\left({1\over \Lambda |t_{f}-t_{i}|}\right)^{{4\over 5} J}.
\end{equation}
The action for the fluctuations of particle
on the effective metric (\ref{eq:effmetric}) is found to be
\begin{equation}
S={1\over 2}\int d\tau 
\left\{ \dot{x}^{2} +\dot{y}_{l}^{2}
+\ell^{2} (x^{2}+{4\over 25}y_{l}^{2}) \right\}
\end{equation}
where $l=1,\ldots 7$. 
The amplitude including the 
contributions from the fluctuations
becomes
\EQ
\left({1\over \Lambda |t_{f}-t_{i}|}
\right)^{{4\over 5}J+{4\over 5}n_{y}+2n_{x}}
,
\label{eq:particleres}
\EN
where $n_{x}$, $n_{y}$ are the occupation numbers for the 
$x$- and $y$- oscillators, respectively. This is consistent
with the result from the supergravity analysis, and also
agrees with the one obtained from the particle on the
true (string-frame) D0-brane background.
However, our analysis in the text clearly shows that 
this kind of effective theory is not meaningful for 
stringy excitations modes as it stands, 
though there may be some different 
ways of extension to stringy excitations.

\end{document}